\documentclass[12pt]{article}
\PassOptionsToPackage{breaklinks}{hyperref}
\pdfoutput=1
\usepackage{amsmath}
\usepackage{amssymb}
\usepackage{cite}
\usepackage[]{graphicx}

\newcommand{\vp}{\varphi}

\newcommand{\tl}{\tilde}

\newcommand{\er}{\eqref}
\begin{document}
\begin{titlepage}
\begin{flushleft}
DESY 08--107\\
 \end{flushleft} \vspace{0.3cm}
\begin{center}
{\Large  On the Advancements of \vspace{0.4cm} \\  Conformal Transformations and their Associated Symmetries
 \vspace{0.5cm} \\
in Geometry and Theoretical Physics\footnote{Invited contribution to the special issue (Sept./Oct.\ 2008) of Annalen der
Physik (Berlin) comme\-mo\-rating Hermann Minkowski's lecture on ``Raum und Zeit'', Sept.\ 1908, in Cologne.}}
 \vspace{1.0cm}\\

  {\large H.A.\
Kastrup\footnote{E-mail: Hans.Kastrup@desy.de}} \vspace{0.4cm}
\\ {DESY, Theory Group \\ Notkestr.\ 85, D-22603 Hamburg
\\Germany} \end{center} 
  \vspace{1cm} \\

{\em \begin{center} Dedicated to the memory of Julius Wess (1934-2007), \\ colleague and friend for many years
\end{center}\vspace{1.0cm} }
 \begin{center}{\bf Abstract}\end{center}
The historical developments of conformal transformations and symmetries are sketched: Their origin from stereographic
 projections of the globe, their blossoming in two dimensions within the field of analytic complex functions, the generic
 role of  transformations by
reciprocal radii in  dimensions higher than two and their linearization in terms of polyspherical coordinates by Darboux, 
 Weyl's attempt to extend General Relativity, the slow rise
of finite dimensional  conformal transformations
in classical  field theories and the problem of their interpretation, then since about 1970 the rapid spread of their
 acceptance for
asymptotic and structural problems in quantum field theories and beyond, 
 up to the
 current $AdS/C\hspace{-1pt}FT$ conjecture. \\ The occasion for the present article: hundred years ago Bateman
 and Cunningham discovered the form invariance of Maxwell's equations for electromagnetism with respect to conformal
space-time transformations. 
\end{titlepage}
 \newpage                  

\tableofcontents 
\section{Introduction}
\subsection{The occasion} Hundred years ago,
on September 21 of 1908, Hermann Minkowski (1864-1909) gave his famous talk on ``Space and Time'' at a congress in
 Cologne \cite{mink1} in which he proposed to unify the traditionally independent notions of space and time in view of 
Einstein's (and Lorentz's) work to a 4-dimensional space-time with a corresponding metric
\begin{equation}
  \label{eq:1}
  (x, x) = (ct)^2 -x^2-y^2-z^2 \equiv (x^0)^2 - (x^1)^2 - (x^2)^2-(x^3)^2, \; \; x= (x^0,x^1,x^2,x^3),
\end{equation}
 to what nowadays is
 called ``Minkowski Space'' $M^4\,$. 

Only a few days later, on October 9, the London Mathematical Society received a paper \cite{bate1} by Harry Bateman
 (1882-1946) in
 which he showed - among others - that the wave equation
 \begin{equation}
   \label{eq:2}
   \frac{1}{c^2}\partial^2_tf(t,\vec{x})-\Delta f(t, \vec{x})=0, \quad\Delta \equiv \partial^2_x +\partial^2_y+
\partial^2_z\,,\,\,\vec{x}=(x,y,z), 
 \end{equation}
is invariant under the (conformal) ``inversion''
\begin{equation}
  \label{eq:3}
  R: \qquad x^{\mu} \to (Rx)^{\mu} \equiv \hat{x}^{\mu} = \frac{x^{\mu}}{(x, x)}\;,\;\mu = 0,1,2,3,
\end{equation}
in the following sense: If $f(x)$ is a solution of Eq.\ \eqref{eq:2}, then
\begin{equation}
  \label{eq:4}
  \hat{f}(x) = \frac{1}{(x, x)}\,f(Rx),\,\,(x, x) \neq 0,
\end{equation}
is a solution of the wave equation, too. Bateman generalized an important result from 1847 by William Thomson
 (Lord Kelvin) (1824-1907) (more details in Subsect.\ 2.3 below) which said: If $h(\vec{x})$ is
 a solution of the
Laplace equation
\begin{equation}
  \label{eq:5}
  \Delta h(\vec{x}) = 0,
\end{equation}
then
\begin{equation}
  \label{eq:6}
  \hat{h}(\vec{x}) = \frac{1}{r}\,h(\vec{x}/r),\; r= (x^2 +y^2 +z^2)^{1/2},
\end{equation}
is a solution, too. In doing so Bateman introduced $w=ict$ and $r^2 = x^2+y^2+z^2+w^2$. In a footnote on pp.\ 75-76 of his
 paper he pointed out that Maxwell's equations, as formulated by H.A.\ Lorentz (1853-1928), take a more symmetrical
 form if the
 variable $ict$ is used. He does not mention Minkowski's earlier introduction of $x_4 = ict$ in his fundamental
treatise on the electrodynamics of moving bodies \cite{mink2}, following the previous work by Lorentz, Poincar\'{e}
 (1854-1912) and Einstein (1879-1955), nor does Bateman mention Einstein's work. But he discusses ``hexaspherical''
coordinates as introduced by Darboux (see Subsect.\ 2.4 below). 

Bateman's paper led, after a few months,  to two more  by himself \cite{bate2,bate3} and one by his colleague
Ebenezer Cunningham (1881-1977) \cite{cunn} in which the form (structure) invariance of Maxwell's electrodynamical 
equations -- including
 non-vanishing charge and current
densities and even special ``ponderable bodies'' -- under conformal space-time
transformations is established, as a generalization of the invariances previously discussed by Lorentz, Einstein
 and Minkowski. 

Bateman's paper is more modern and more elegant in that he uses efficiently a precursor of differential forms
 (from 1-forms up to 4-forms) for his arguments. 

In both papers there is no discussion of  possible connections of the newly discovered additional
form invariance of Maxwell's equation to new conservation laws. Here  the remark is important that form invariance of
 differential equations with respect to
certain transformations in general leads to new solutions (see, e.g.\ Eqs.\ \er{eq:5} and \er{eq:6}), but
 not necessarily to new conservation laws! See Sect.\ 4 for
more details.

Bateman also speculated \cite{bate2} that the conformal transformations may be related to accelerated motions, an issue we
shall encounter again below (Subsect.\ 5.2). 

The ``correlations'' between the two authors of the papers \cite{bate2} and \cite{cunn} are not obvious, but the initiative
appears to have been on Bateman's side: In a footnote on the first page of his paper Cunningham says: ``This paper
 contains in
an abbreviated form the chief parts of the work contributed by the Author to a joint paper by Mr.\ Bateman and himself
read at the meeting held on February 11th, 1909, and also the work of the paper by the author read at the meeting held on
March 11th, 1909.'' And in a footnote on the third page Cunningham remarks: ``This was pointed out to me by Mr.\ Bateman,
a conversation with whom suggested the present investigation.'' Here Cunningham is refering to  invariance of
the wave equation under the transformation by reciprocal radii Bateman had investigated before \cite{bate1}. In the
essential part II of his paper Cunningham first gives the transformation formulae for the electric and magnetic {\em fields}
with respect to the inversion \er{eq:3} and says in a footnote on p.\ 89 of Ref.\ \cite{cunn} that the
 corresponding formulae for the
scalar and vector {\em potentials} were suggested to him by Bateman. 

 Bateman does not mention a joint paper with Cunningham which, as far as I know, was never published.
He also  read his paper \cite{bate2} at the meeting of the Mathematical Society on March 11th. On the second page
of his article \cite{bate2} he says: ``I have great pleasure in thanking Mr.\ E.\ Cunningham for the stimulus which he
gave to this research by the discovery of the formulae of transformation in the case of an inversion in the
 four-dimensional space.'' And in his third
paper \cite{bate3} when he discusses transformation by reciprocal radii Bateman says: ``Cunningham \cite{cunn} has shown
that any electrodynamical field may be transformed into another by means of this transformation.'' 

So it is not clear who of the two -- after Bateman's first paper \cite{bate1} on the wave equation -- had the idea or
 suggested to look for conformal invariance of Maxwell's electrodynamics, and why the initial joint paper was not
 published. Perhaps the archives of the London Mathematical Society can shed more light on this. From the publications
one may conclude that Bateman found the transformations with respect to the inversion \er{eq:3} for the potentials and
Cunningham -- independently -- those for the fields!

{\em Those papers by Bateman and Cunningham were the beginning of discussing and applying conformal transformations
 in modern physical field theories. But it took  more than
50 years till the physical meaning of those conformal transformations became finally clarified and its general role in
 theoretical physics fully established.} From about 1965/70 on conformal symmetries have been creatively and
 successfully exploited for
 many physical systems or their more or less strong idealizations. The  emphasis of these notes -- which are not complete
at all -- will be on  different stages
  till about 1970 of that period and
 they will mention more recent developments more superficially, because there are many modern reviews on the topics
of those activities. 

\subsection{The issue} Conformal transformations of geometrical spaces with a metric may appear in two different ways:
\subsubsection{Conformal mappings as point transformations}

 Let $\mathcal{M}^n, n \geq 2,$ be an n-dimensional Riemannian or pseudo-Riemannian manifold  with local
 coordinates
 $x=(x^1,\ldots,x^n)$
 and endowed with a  (pseudo)-Riemannian non-degenerate metric
\begin{equation}
  \label{eq:7}
 g_x = \sum_{\mu,\nu =1}^n g_{\mu \nu}(x)\,dx^{\mu}\otimes dx^{\nu},
\end{equation}
i.e.\ if 
\begin{equation}
  \label{eq:19}
  a= \sum_{\mu=1}^n a^{\mu}(x)\partial_{\mu}, \qquad b=\sum_{\nu=1}^nb^{\nu}(x)\partial_{\nu},
\end{equation}
are two tangent vectors at the point $x$, then they have the scalar product
\begin{equation}
  \label{eq:20}
  g_x(a,b) = \sum_{\mu,\nu =1}^n g_{\mu \nu}(x)\,a^{\mu}\, b^{\nu},
\end{equation}
and the cosine of the angle between them is given by
\begin{equation}
  \label{eq:21}
  \frac{g_x(a,b)}{\sqrt{g_x(a,a)}\,\sqrt{g_x(b,b)}}\,.
\end{equation}
Let $\hat{\mathcal{M}}^n$ be a second corresponding manifold with local coordinates $\hat{x}^{\mu}$ and metric
$\hat{g}_{\hat{x}}$.
Then a mapping
\begin{equation}
  \label{eq:8}
  x \in G \subset \mathcal{M}^n \to \hat{x} \in \hat{G}  \subset \hat{\mathcal{M}}^n
\end{equation}
is said to be {\em conformal} if
\begin{equation}
  \label{eq:9}
  \hat{g}_{\hat{x}} = C(x)\,g_x, \quad C(x) \neq 0, \infty,
\end{equation}
where the function $C(x)$ depends on the mapping.
The last equation means that the angle between two smooth curves which meet at x is the same as the angle between the
 corresponding
image curves meeting at the image point $\hat{x}$. Note that the mapping \eqref{eq:8} does not have to be defined on
 the whole $\mathcal{M}^n$. 
\\ \\
{\em  Two important examples:} 
\\ \\
{\bf I. Transformation by reciprocal radii} \hfill \\

   For the  inversion \eqref{eq:3} (mapping  by ``reciprocal radii'' of
 the Minkowski space
 into itself)  we have
\begin{equation}
  \label{eq:11}
  (\hat{x},\hat{y})= \frac{(x,y)}{(x,x)\,(y,y)},
\end{equation}
and
\begin{eqnarray}
  \label{eq:12}
  \hat{g}_{\hat{x}}& \equiv & (d\hat{x}^0)^2-  (d\hat{x}^1)^2- (d\hat{x}^2)^2- (d\hat{x}^3)^2  = \frac{1}{(x,x)^2}\,g_x, 
\\   g_x &=& (dx^0)^2 -(dx^1)^2-(dx^2)^2-(dx^3)^2\,. \nonumber
\end{eqnarray}

These equations show again that the mapping \eqref{eq:3} is not defined on the light cone $(x,x) = 0$. We shall later see
how this problem can be cured by adding points at infinity, i.e.\ by extending the domain of definition for the mapping
\eqref{eq:3}. 

It will be discussed in the next Sect.\ that there is an important  qualitative difference as to conformal mappings
of Euclidean or pseudo-Euclidean spaces $\mathbb{R}^n$ with a metric
\begin{equation}
  \label{eq:22}
  (x,x)= \sum_{\mu,\nu=1 }^n \eta_{\mu \nu}\,x^{\mu}x^{\nu}, \quad \eta_{\mu \nu} = \pm \delta_{\mu \nu},
\end{equation}
 for $n=2$ and for
$ n > 2\,$: \\ For $n=2$ any holomorphic or meromorphic function 
\begin{equation}
  \label{eq:13}
  w=u+i\,v = f(z),\,z=x+iy
\end{equation}
provides a conformal map of regions of the complex plane:
\begin{equation}
  \label{eq:14}
  (du)^2 + (dv)^2 = |f'(z)|^2\,[(dx)^2 + (dy)^2] .
\end{equation}
Here it is assumed that $f'(z) = df/dz $ does not vanish at $z$ and that the Cauchy-Riemann eqs.\ hold (see Eq.\
 \eqref{eq:17} below). A map by such a holomorphic or meromorphic function also preserves the orientation
of the angle. On the other hand, a corresponding antiholomorphic function $g(z^*),\,z^* = x-iy,$ does preserve
 angles, too, but reverses their orientations. \\ One here can, of course, go beyond the complex plane to Riemann surfaces
with more complicated structures.

For $n > 2$, however,  conformal mappings constitute ``merely'' a [$(n+1)(n+2)/2$]-dimensional Lie transformation group
 which may be generated by the inversion
\begin{equation}
  \label{eq:15}
   R:~ x^{\mu} \to (Rx)^{\mu} \equiv \hat{x}^{\mu} = \frac{x^{\mu}}{(x, x)}, \;\;\mu = 1,\ldots,n >2,\,\, R^2 = 1,
\end{equation}
and the translations
\begin{equation}
  \label{eq:16}
  T_n(b): \qquad x^{\mu} \to \hat{x}^{\mu} = x^{\mu} + b^{\mu},\,b^{\mu} \in \mathbb{R},\,\, \mu =1,\ldots,n.
\end{equation} 
{\bf II.\  Stereographic projections} \hfill \\

A historically very important example is the stereographic projection of the surface $S^2$ of a sphere 
 with radius $a$ in $\mathbb{R}^3$ onto the plane (see Fig.~\ref{fig:stereographic}):
 \begin{figure}[h]
   \centering
   \includegraphics[page=1]{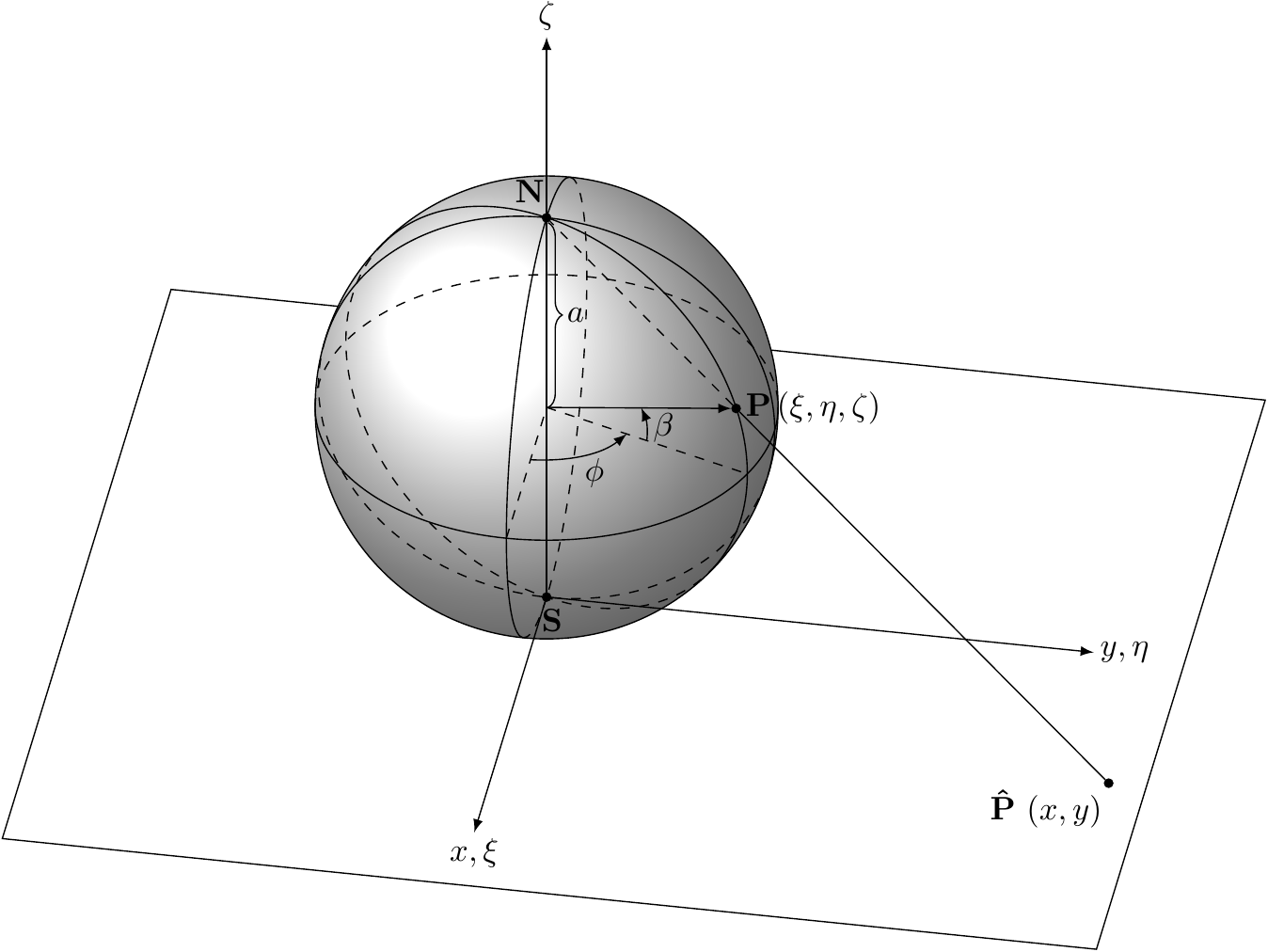}
\caption{\protect{{\bf Stereographic projection:}} The points $P$ on the surface of a sphere with radius $a$ are mapped onto
points $\hat{P}$ in the plane -- and vice versa -- by drawing a straight line from the north pole $N$ of the sphere
through $P$ towards $\hat{P}$. The mapping is conformal and arbitrary circles on the sphere are mapped onto circles or
straight lines in the plane.}
   \label{fig:stereographic}
 \end{figure}

 Let the south pole of the sphere coincide with the origin $(x,y) = (0,0)$ of the plane and its north pole with
the point $(\xi=0=x,\eta=0=y, \zeta =2a) \in \mathbb{R}^3$. The projection is implemented by connecting the north pole with 
 points $\hat{P}(x,y)$ in the
plane by  straight lines which intersect the surface of the sphere in the points $P(\xi,\eta,\zeta) $ with 
$\xi^2 +\eta^2 +(\zeta-a)^2 =a^2$. In this way the point $P(\xi,\eta,\zeta)$ of the sphere is mapped into the point
 $\hat{P}(x,y)$
of the plane. 

 Analytically the mapping is given by
\begin{equation}
  \label{eq:23}
  x = \frac{2a\,\xi}{2a-\zeta}, \quad y=\frac{2a\,\eta}{2a-\zeta}, \quad\xi^2+\eta^2+\zeta^2-2a\,\zeta =0,
\end{equation}
with the inverse map
\begin{equation}
  \label{eq:24}
  \xi = \frac{4a^2\,x}{4a^2+x^2+y^2}, \quad \eta=\frac{4a^2\,y}{4a^2+x^2+y^2}, \quad\zeta =\frac{2a\,(x^2+
y^2)}{4a^2+x^2+y^2}.
\end{equation} Note that the north pole of the sphere is mapped to ``infinity'' of the plane which has to added as a
 ``point''
in order to make the mapping one-to-one!

Parametrizing the spherical surface by an azimutal angle $\phi$ (``longitude'') in its equatorial plane $\zeta=a$ parallel
 to the $(x,y)$-plane, with
the initial meridian ($\phi = 0$) given by the plane $y=\eta=0$, and the angle $\beta$ (``latitude'') between that
 plane and 
the position vector of the
point $(\xi,\eta,\zeta)$, with respect to the centre of the sphere, with $\beta$ positive on the northern half
 and negative on
the southern half. We then have
\begin{equation}
  \label{eq:25}
  \xi = a\,\cos \phi\,\cos\beta, \quad \eta = a\,\sin\phi\,\cos\beta, \quad \zeta-a = a\,\sin\beta,
\end{equation}
and
\begin{equation}
  \label{eq:26}
  x=\frac{2a\,\cos\phi\,\cos\beta}{1-\sin\beta}, \quad  y=\frac{2a\,\sin\phi\,\cos\beta}{1-\sin\beta}, \quad 
\frac{\cos\beta}{1-\sin\beta}= \tan(\beta/2 +\pi/4).
\end{equation}
The last equations imply
\begin{equation}
  \label{eq:27}
  g_{(x,y)} = (dx)^2+(dy)^2 = \frac{4}{(1-\sin\beta)^2}\; g_{(\phi,\beta)}, \quad g_{(\phi,\beta)}=a^2[\cos^2\beta\,(d\phi)^2 
+(d\beta)^2].
\end{equation}
Here $g_{(\phi,\beta)}$ is the standard metric on a sphere of radius $a$. Eq.\ \eqref{eq:27} shows that the stereographic
pro\-jection \eqref{eq:26} is a conformal one, with the least distortions of lengths from around the south pole
($\sin \beta \geq -1$).

Besides being conformal,  the stereographic projection given by the Eqs.\ \eqref{eq:23} and \eqref{eq:24} has the
second important property that circles on the sphere are mapped onto the circles on the plane (where straight lines are
 interpreted
as circles of infinite radii) and vice versa. This may be seen as follows: Any circle on the sphere can be generated 
by the intersection of the sphere with a plane
\begin{equation}
  \label{eq:65}
  c_1\,\xi + c_2\, \eta + c_3\,\zeta + c_0 = 0.
\end{equation}
Inserting the relations \eqref{eq:24} with $2a =1$ into this equation yields
\begin{equation}
  \label{eq:66}
  (c_0 + c_3)(x^2 + y^2) + c_1\,x + c_2 y + c_0 = 0,
\end{equation}
which for $c_0 + c_3 \neq 0$ describes  the circle
\begin{equation}
  \label{eq:67}
  (x + \tilde{c}_1 /2)^2 + (y+\tilde{c}_2 /2)^2 = \rho^2, \quad  \tilde{c}_j = \frac{c_j}{c_0+c_3},\, j=0,1,2\,;
 \quad \rho^2 = ( \tilde{c}_1^2 + \tilde{c}_2^2)/4 - \tilde{c}_0.
\end{equation}
The coefficients $c_j$ in Eq.\ \eqref{eq:65}  have to be such that the plane actually intersects or touches the plane. This
means that $\rho^2 \geq 0$ in Eq.\ \eqref{eq:67}.

If $c_0+ c_3 = 0, \, c_0 \neq 0$, then the Eqs.\ \eqref{eq:65} and \eqref{eq:66} can be reduced to
\begin{equation}
  \label{eq:69}
  \hat{c}_1\,\xi + \hat{c}_2\, \eta - \zeta + 1 = 0 
\end{equation} and
\begin{equation}
  \label{eq:70}
   \hat{c}_1\,x + \hat{c}_2\, y + 1 = 0.
\end{equation}
Here the plane \eqref{eq:69} passes through the north pole $(0,0,1)$ and the image of the associated circle on the sphere
is the  straight line \eqref{eq:70}. 

If $c_3 = c_0 = 0$ then the plane \eqref{eq:65}  contains a meridian and
Eq.\ \eqref{eq:66} becomes a straight line through the origin.

On the other hand the inverse image of the circle
\begin{equation}
  \label{eq:71}
  (x-\alpha)^2 + (y-\beta)^2 = \rho^2
\end{equation}
is, according to the Eqs.\ \eqref{eq:23}, associated with the plane
\begin{equation}
  \label{eq:72}
  2\alpha\,\xi + 2\beta\,\eta + (\alpha^2 + \beta^2 - \rho^2 -1)\,\zeta + \rho^2 - \alpha^2 - \beta^2 = 0,
\end{equation}
where the relation $\xi^2 + \eta^2 = \zeta\,(1-\zeta)$ has been used. \indent

As the stereographic projection plays a very crucial role in the long history of conformal transformations, up to the newest
developments, a few historical remarks are appropriate: 

 The early interest in stereographic projections was strongly influenced by its applications to the construction of
the astrolabe -- also called planisphaerium --, an important (nautical) instrument \cite{astr} which
 used a  stereographic projection for describing properties of the celestial (half-) sphere in a plane. It  may have
 been known already at the time of Hipparchos (ca.\ 185 -- ca.\ 120 B.C.) \cite{neu1}. It was definitely used for that
 purpose by Claudius Ptolemaeus (after 80 -- about 160 A.D.) \cite{ptol}. Ptolemaeus knew that circles are mapped onto
circles or straight lines by that projection, but it is not clear whether he knew that {\em any} circle on the sphere
 is mapped onto a circle 
or a straight line. That  property was proven
by the astronomer and engineer  Al-Fargh\={a}n\={\i} (who lived in Bagdad and Cairo in the first half of the 9th century)
 \cite{far} and independently briefly after 1200 by the
European mathematician ``Jordanus de Nemore'', the identity of which appears to be unclear \cite{nem}. 

That the stereographic projection is also conformal was explicitly realized considerably later: In his book on the
``Astrolabium'' from 1593 the mathematician and Jesuit Christopher Clavius (1537-1612) showed how to determine the
angle at the intersection of two great circles on the sphere by merely measuring the corresponding angle of their images
on the plane \cite{clav}. This is equivalent to the assertion that the projection is conformal \cite{hall}. 

Then there is Thomas Harriot (1560-1621) who about the same time also showed  -- in unpublished and undated notes -- that 
the stereographic projection is conformal. Several remarkable mathematical, cartographical and physical discoveries of this
ingenious nautical adviser of Sir Walter Raleigh (ca.\ 1552-1618) were rediscovered and published between 1950 and 1980
 \cite{harr1}. During his lifetime Harriot published none of his mathematical insights and physical experiments
\cite{harr2}. His notes on the conformality of stereographic projections have been dated (not conclusively) 
between 1594 and 1613/14 \cite{harr3}, the latter date appearing more likely. So in principle Harriot could have known
Clavius' Astrolabium \cite{harr4}. 

 In 1696 Edmond Halley (1656-1742) presented a paper to the Royal Society of
London in which he proves the stereographic projection to be conformal, saying that Abraham de Moivre (1667-1754) told
him the result and that Robert Hooke (1635-1703) had presented it before to the Royal Society, but that the present proof
was his own \cite{hally}. 
\subsubsection{Weyl's geometrical gauge transformation}
 A second way of implementing a conformal transformation for a Riemannian or pseudo-Riemannian mani\-fold is
 the possibility 
of  merely multiplying the metric form \eqref{eq:7} by a non-vanishing positive smooth function $\omega(x) > 0$:
\begin{equation}
  \label{eq:18}
  g_x \to \hat{g}_x = \omega(x) \, g_x.
\end{equation}
More details for this type of conformal transformations, introduced by Hermann Weyl (1885-1955), are discussed below
 (Sect.\ 3).

The Eqs.\ \eqref{eq:9} and \eqref{eq:18} show that the corresponding conformal mappings change the length scales of the
systems involved. As many {\em physical} systems have inherent fixed lengths (e.g.\ Compton wave lengths (masses) of
 particles, coupling constants
with non-vanishing length dimensions etc.), applying the above conformal transformations to them in many cases 
cannot lead to genuine
symmetry operations, like, e.g.\ translations or rotations. As discussed in more detail below, these limitations are
one of the reasons for the slow advance of conformal symmetries in physics! 

 Here it is very important to emphasize the difference between  transformations which merely change the coordinate frame 
and  the analytical description of a system and those mappings where the coordinate system is kept fixed:
 in the former case the system under consideration, e.g.\
 a hydrogen atom with its discrete and
continuous spectrum, remains the same, only the description changes; here one may choose any macroscopic unit of energy or
an equivalent unit of length in order to describe the system. However, in the case of mappings one 
 asks whether there are other
 systems than the given one which can be considered as images of that initial system for the  mapping under consideration.
 But now, in the case of dilatations, there
is no continuous set of hydrogen atoms the energy spectra of which  differ from the the original one by  arbitrary
scale transformations! For the existence of conservation laws the invariance with respect to mappings is crucial
 (see Sect.\ 4 below). These two types of transformations more recently have also been called ``passive'' and ``active''
 ones \cite{wigh}.
\section{Conformal mappings till the end of the 19th century}
\subsection{Conformal mappings of 2-dimensional surfaces}

\begin{figure}[h]
\centering
\includegraphics[page=2]{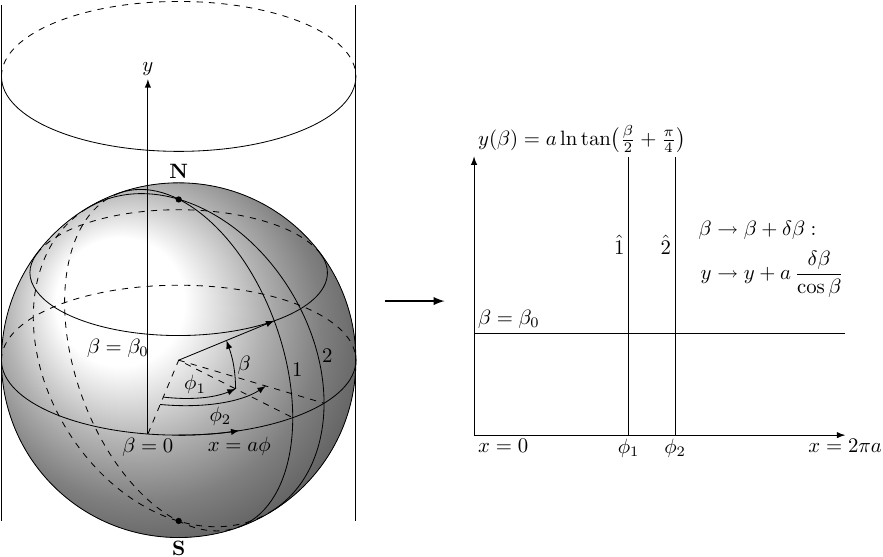}
\caption{\protect{{\bf Mercator projection:}} The points (longitude $\phi$, latitude $\beta$) on the surface of a sphere
 with radius $a$ 
are mapped on the mantle of a cylinder which touches the equator and is then unrolled onto the plane. The circles of
fixed latitude $\beta$ are mapped onto straight lines parallel to the $x$-axis, different meridians are mapped onto
 parallel lines along the $y$-axis, itself  parallel to the cylinder axis. The mapping is characterized by the
property that an increase $\delta \beta$ in latitude implies an increase $\delta y = a\,\delta \beta/\cos \beta$ on
the cylinder mantle. This makes it a conformal one.}
\label{fig:mercator} \end{figure}

With the realization that the earth is indeed a sphere and the discoveries of faraway continents the need for maps of
 its surface became urgent, especially for
ship navigation. Very important progress in cartography \cite{cart1}  was made by Gerhardus Mercator (1512-1594),
 particularly with his world map
from 1569 for which he employed a conformal ``cylindrical'' projection \cite{cart2}, now named after him \cite{merc}.

 Here
the meridians are projected onto parallel lines on the mantle of  a cylinder
 which touches the equator of the sphere  and which is unwrapped onto a plane afterwards (see Fig.~\ref{fig:mercator}):
  Let $\phi$ and $\beta$
 have the same meaning
as in example 2 of Subsect.\ 1.2 above (longitude and latitude). If $a$ again is the radius of the sphere, $x$ the
 coordinate around the cylinder where it touches the equator, then $x = a\, \phi$ . The orthogonal $y$-axis on the
 mantle of the cylinder, parallel to its axis, meets the equator at $\phi =0=x,\, \beta = 0 =y$. The mapping of the
 meridians onto parallels of the $y$-axis on the mantle is
determined by the requirement that $ \cos \beta\, \delta y = a\, \delta \beta$ for a small increment $\delta \beta$.

Thus the Mercator map is characterized by
\begin{equation}
  \label{eq:29}
  \delta x = a\, \delta \phi, \quad \delta y = \frac{a}{\cos \beta}\, \delta \beta,
\end{equation}
yielding
\begin{equation}
  \label{eq:30}
  (dx)^2 + (dy)^2 = \frac{1}{\cos^2 \beta}\, g_{(\phi,\beta)}, \quad g_{(\phi,\beta)}=a^2[\cos^2\beta\,(d\phi)^2 
+(d\beta)^2],
\end{equation}
which shows the mapping to be conformal, with strong length distortions near the north pole! (Integrating the
 differential equation $dy/d\beta = 1/\cos \beta$ gives $y(\beta) = \ln \tan(\beta/2 + \pi/4)$ with $y(\beta = 0)=0$.)

 A first pioneering explicit mathematical ``differential'' analysis of Mercator's projection, the stereographic
 projection and the more general problem of mapping the surface
of a sphere onto a plane was published  in 1772 \cite{lamb} by Johann Heinrich Lambert (1728-1777): Lambert posed the
problem which ``global'' projections of a spherical surface onto the plane are compatible with local (infinitesimal)
 requirements
like angle-preserving or area-preserving, noting that both properties cannot be realized simultaneously!
 He showed that his differential conditions for angle preservation are fullfilled
by Mercator's and the stereographic projection. In addition he presented a new ``conical''
 -- also conformal -- solution,  still known and used as ``Lambert's projection'' \cite{lamb2}. 

The term ``stereographic projection'' was introduced by the Belgian Jesuit and mathematician (of Spanish origin)
 Fran\c{c}ois d'Aiguillon (1567-1617)
in 1613 in the sixth and last part -- dealing with projections (``Opticorum liber sextus de proiectionibus'') --
 of his book on optics \cite{aig} which became also well-known for its engravings
 by the painter Peter Paul Rubens (1577-1640) at the beginning of each of the six parts and on the title page \cite{aig2}!
When introducing the 3 types of projections he is going to discuss (orthographic, stereographic and scenographic ones) 
d'Aiguillon
says almost jokingly \cite{aig3}: ``... Second,'' [projection] ``from a point of contact'' [on the surface of the sphere],
`` which not
 improperly
could be called stereographic: a term that might come into use freely, as long as no better one occurs, if you, Reader,
allows for it''.  The readers did allow for it!

 Only three years after the
publication of Lambert's  work  Leonhard Euler (1707-1783) in 1775 presented three communications to the Academy of 
St.\ Petersburg (Russia) on
 problems 
concerning (cartographical) mappings from the surface of a sphere onto a plane, the first two being mainly mathematical.
 The papers were published in 1777
 \cite{eul}. Euler approached the problem Lambert had posed from a more general point of view by looking for a larger class
of solutions of the differential equations by using methodes he had employed previously in 1769 \cite{eul2}. In his
``Hypothesis 2'' (first communication) Euler formulated the differential equations for the condition that small parts on
 the earth are mapped on
similar figures on the plane (``Qua regiones minimae in Terra per similes figuras in plano exhibentur''), i.e.\ the mapping
should be conformal. For obtaining the general solution of those differential equations Euler  
used complex coordinates $z=x+iy$ in the plane. This appears to be the first time that such a use of complex variables was
made \cite{kom}. Euler further  observed (second  communication)
 that the mapping
\begin{equation}
  \label{eq:10}
  z \to \frac{a \,z+b}{c\,z+ d}, \quad z= x+iy,
\end{equation}
which connects the different projections in the plane is a conformal one! Euler does not mention Lambert's work
 nor did Lambert mention Euler's earlier paper from 1769 on the construction of a family of curves which are orthogonal 
to the curves of a given family \cite{eul2}.
As Euler had supported a position for Lambert in Berlin in 1764 before he - Euler - left for St.\ Petersburg in 1766, this
mutual silence is somewhat surprising. 

 Lambert  mentions in his article that he informed Joseph-Louis de Lagrange 
(1736-1813) about the cartographical problems he was investigating. In 1779 Lagrange, who was in Berlin since 1766 as
president of the Academy,
 presented two
longer Memoires on the construction of geographical maps to the Berlin Academy of Sciences \cite{lagr}.
 Lagrange says that he wants to generalize the
work of Lambert and Euler and look for all projections which map circles on the sphere onto circles in the plane. 

The more general problem of mapping a 2-dimensional (simply-connected) surface onto another one while preserving angles
 locally was finally solved
completely in 1822 by Carl Friedrich Gauss (1777-1855) in a very elegantly written paper \cite{gauss1} in which he showed
 that the
general solution is given by  functions of  complex numbers $q+ip$ or $q-ip$. As he assumes differentiability of the
functions with respect to their complex arguments he - implicitly - assumes the validity of the Cauchy-Riemann differential
equations \eqref{eq:17}! 

 Gauss  does not use the  term ``conform'' for the
mapping in his 1822 paper, but he
introduces it in  a later one from 1844 \cite{gauss2}. The expression ``conformal projection ''
 appears for the first time as ``proiectio conformis''
in an  article written in Latin and presented in 1788 to the St. Petersburg (Russia) Academy of Sciences by the German-born
 astronomer and mathematician
Friedrich Theodor Schubert (1758-1825)  \cite{schub}. It was probably the authority of
Gauss which finally made that term ``canonical''!

 The development was brought to a certain culmination by
Gauss' student Georg Friedrich Bernhard Riemann (1826-1866) who in his Ph.D.\ Thesis \cite{Rie} from 1851
 emphasized the important
 difference between global and local properties of 2-dimensional surfaces described by functions of complex variables 
and who formulated his famous version of Gauss' result
(he quotes Gauss' article from 1822 at the beginning of his paper; except for a mentioning of Gauss' paper from 1827
 \cite{gau3} at the
end,  this is the only reference Riemann gives!),
namely that every simply-connected region of the complex plane can be mapped (conformally) into the interior of the
unit circle $|z| <1$ by a holomorphic function \cite{ahl}.

After writing down the conditions (Cauchy -- Riemann equations \cite{cau})
\begin{equation}
  \label{eq:17}
  \frac{\partial u}{\partial x} = \frac{\partial v}{\partial y}, \quad \frac{\partial u}{\partial y} =
- \frac{\partial v}{\partial x},
\end{equation}
for the uniqueness of differentiating a complex function $ w = u+iv = f(z = x+iy)$ with respect to $z$, Riemann notes that
they imply the second order [Laplace] equations
\begin{equation}
  \label{eq:28}
  \frac{\partial^2 u}{\partial x^2} + \frac{\partial^2 u}{\partial y^2}= 0, \quad 
\frac{\partial^2 v}{\partial x^2} + \frac{\partial^2 v}{\partial y^2}= 0.
\end{equation}
Conformal mappings, of course, still play a central role for finding solutions of
 2-dimensional Laplace equations which obey given boundary conditions in a vast variety of applications \cite{pot1}.
 This brings us -- slowly -- back to 
the history of Eqs.\ \eqref{eq:5} and \eqref{eq:6}:
\subsection{On circles, spheres, straight lines and reciprocal radii}
The history of conformal mappings described in the last Subsect.\ represents the beginning of modern differential geometry
 -- strongly induced by new cartographical challenges -- which culminated in Gauss' famous paper from 1827  on
the general theory of curved surfaces \cite{gau3}. Another pillar of that development was the influential work of the French
mathematician Gaspard Monge (1746-1818), especially by his book on the application of analysis to geometry \cite{mon1}.
Through his students he also influenced  the subject to be discussed now:

 About the  time of 1825 several of those mathematicians which were more interested in the 
global purely {\em geometrical} relationships between circles and lines, spheres and planes and more complicated geometrical
objects (so--called ``synthetic'' or ``descriptive geometry'' \cite{syn,hach}, as contrasted to the -- more
 modern -- analytical geometry) discovered the mapping by reciprocal radii (then also called ``inversion''): 

 A seemingly thorough, balanced and informative 
 account of that period and the questions of priorities involved  was
given in 1933 by Patterson \cite{pat}. He missed, however, a crucial paper from 1820 by a 22 years old self-educated
 mathematician, who died only 5 years later.
In view of the general scope of Patterson's work I can confine myself to a few additional illustrating, but crucial, 
remarks on the rather complicated and bewildering beginning of the concept ``transformation by reciprocal radii'':

There is the conjecture that the Swiss mathematician Jakob Steiner (1796-1863) was the first to know the mapping around
 the end of 1823
or the beginning of 1824. The corresponding notes and a long manuscript were found long after his death and not
 known when his
collected papers where published \cite{stein1}.  In a first publication of notes from Steiner's literary estate in 1913 by
B\"utzberger \cite{bue}
it was argued that Steiner knew the inversion at least in February 1824 and that he was the first one.
 In 1931 a long manuscript by Steiner
 on circles and spheres
from 1825/1826 was finally published \cite{stein2} which also shows Steiner's vast knowledge of the subject. \\
However, the editors of that
manuscript, Fueter and Gonseth, say in their introduction that in January 1824 Steiner made extensive excerpts
 from a long paper
 by the young French mathematician J.B. Durrande (1798-1825), published in July
1820 \cite{durr1}.\ From that they draw the totally unconvincing conlusion that Steiner knew already what he extracted
from the journal!  A look at Durrande's paper shows immediately that he definitely deserves
the credit for priority \cite{durr6}! Not much is known about this self-educated mathematician who died at the age of
27  years: From March 1815 till October 1825 twenty eight papers by Durrande were published in  Gergonne's
Annales \cite{gerg1}, the last one after his death \cite{durr2}. In a footnote on the title page of Durrande's first
paper Gergonne remarks that the author is a 17 years old geometer who learnt mathematics only with the help of
 books \cite{durr3}.

Many of Durrande's contributions present solutions of problems which had been posed in the Journal previously, most
of them by Gergonne himself. The important paper of July 1820 originated, however, from Durrande's own conceptions. In it
he appears with the title `` professeur de math\'ematiques, sp\'eciales et de physique au coll\'{e}ge
royal de Cahors'' and in his second
last paper from November 1824 \cite{durr4} as `` professeur de physique au coll\'{e}ge royal de
Marseille'' \cite{durr5}. At the end of that paper and in his very last one, Ref.\ \cite{durr2}, Durrande again used
 the inversion, he had introduced before in his important paper from 1820. 

The Annales de Gergonne were full of articles dealing with related geometrical problems. Steiner himself published 8 papers
in volumes 18 (1827/28) and 19 (1828/29) of that journal. 

Around 1825 the two  Belgian mathematicians and friends Germinal Pierre Dandelin (1794-1847) and Lambert Adolphe
 Jacques Quetelet 
(1796-1874) were investigating very similar problems, presenting their results to the L'Acad\'emie Royal des Sciences
et Belles-Lettres de Bruxelles \cite{patt2}. At the end of a paper by Dandelin, presented on June 4 of 1825, there is
the main formula of inversion \cite{dand1} (see Eq.\ \eqref{eq:32} below) and at the end of a longer paper
 by Quetelet \cite{quet}, presented
 on November 5 of the
same year, a 3-page note is appended which contains -- probably for the first time in ``analytical'' form -- the
 transformation formula 
\begin{equation}
  \label{eq:31}
  \hat{x}= \frac{r_0^2\,x}{x^2 + y^2},\quad \hat{y} = \frac{r_0^2\,y}{x^2 + y^2},
\end{equation}
for an inversion on a circle with radius $r_0$.

The transformation \eqref{eq:31} is also mentioned by Julius Pl\"{u}cker (1801-1868) in the first volume of his textbook
from 1828 \cite{plue}.

Other important early contributions to the subject (mostly ignored in the literature) are those of the Italian mathematician
Giusto Bellavitis (1803-1880) in 1836 and 1838 \cite{bell}. \\
Now back to the mathematics \cite{klei,enc}! \\
\begin{figure}[h]
  \centering
  \includegraphics[page=3]{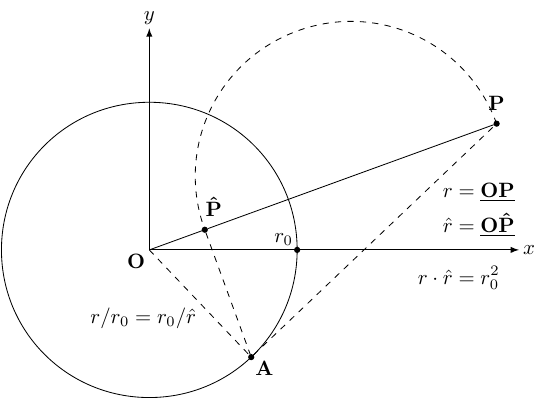}
 \caption{\protect{{\bf Inversion on a circle with radius $\mathbf{r_0}$:}} A point $P$ outside the circle with distance r
 from the center $O$ 
 is mapped onto a point $\hat{P}$ on the line $\underline{OP}$ with distance $\hat{r} = r^2_0/r$. A line from
$P$ tangent to the circle at $A$ generates several similar rectangular tri\-angles corresponding sides of
 which obey $r/r_0 = r_0/
\hat{r}$. A circle through $P$ and $\hat{P}$ with its origin on $\underline{\hat{P}P}$ is orthogonal to the original one.}
  \label{fig:inversion}
\end{figure}

The basic geometrical idea is the following (see Fig.~\ref{fig:inversion}): Given a circle with radius
 $r_0$ and
 origin $O$
in the plane, draw a line from the origin to a point $P$ {\em outside} the circle with a distance $r$ from the origin $O$
away. If a point $\hat{P}$ on the same line {\em inside} the circle and with the distance $\hat{r}$ form the origin
obeys the relation
\begin{equation}
  \label{eq:32}
  r\,\hat{r} = r_0^2,
\end{equation}
then the point $\hat{P}$ is called ``inverse'' to $P$ and vice versa.

 The last equation obviously is a consequence of Eqs. 
\eqref{eq:31}. If now $P$ traces out a curve then $\hat{P}$ describes an
``inverse'' curve, e.g.\ circles are mapped onto circles (see below).  $P$ and $\hat{P}$ were also called ``conjugate''.
 Such points have many interesting geometrical
properties, e.g.\ if one draws a circle through the points $P$ and $\hat{P}$, with its origin on the line connecting the
two inverse points, then the new circle is  orthogonal to the old one! For many more interesting properties of such
systems see the textbooks mentioned in Refs.\ \cite{klei,enc}. 

If
\begin{equation}
  \label{eq:33}
B \equiv  (x-\alpha)^2 + (y-\beta)^2 - \rho^2 = x^2 + y^2 -2\alpha \,x - 2\beta \, y +C = 0,\,\,\, C= \alpha^2 +
 \beta^2 - \rho^2,
\end{equation}
is any circle  in the plane with radius $\rho$, then the ``inverse'' circle $\hat{B}$ generated by the transformation 
\eqref{eq:31} has the constants
\begin{equation}
  \label{eq:34}
  \hat{\alpha} = \frac{r_0^2\,\alpha}{C},\quad \hat{\beta} = \frac{r_0^2\,\beta}{C},\quad \hat{C} =
\frac{r_0^4}{C}, \quad \hat{\rho}^2 = \frac{r_0^4}{C^2}\,\rho^2.
\end{equation}
Eqs.\ \eqref{eq:31} show that the ``inverse'' of the origin $(x=0, y=0)$ is infinity! If a circle \eqref{eq:33} passes
through the origin, then $C=0$ and it follows from the last of the Eqs.\ \eqref{eq:34} that $\hat{\rho} = \infty$, i.e.\
the image of a circle passing through the origin is one with an infinite radius, that is a straight line! It follows from
the Eqs.\ \eqref{eq:31}, \eqref{eq:33} and $C=0$ that this image straight line obeys the equation
\begin{equation}
  \label{eq:35}
  \alpha\,\hat{x} + \beta\,\hat{y} -r_0^2/2 = 0.
\end{equation}
On the other hand, a straight line given by
\begin{equation}
  \label{eq:36}
  b_1\,x + b_2\, y + g = 0 ,\;\; g \neq 0,
\end{equation}
is mapped onto the circle
\begin{equation}
  \label{eq:37}
  (\hat{x} +r_0^2\,b_1/2g)^2 + (\hat{y} + r_0^2\,b_2/2g)^2 = (r_0^2\,b_1/2g)^2 + (r_0^2\,b_2/2g)^2.
\end{equation}
In order to have the mapping \eqref{eq:31} one-to-one  one has to add a {\em point} at infinity (not a straight line
 as in projective geometry!). The situation is completely the same as in the case of stereographic projections discussed
in example II of Subsect.\ 1.2.1 above. Thus, the set (totality) of circles and straight lines in the plane is
 mapped onto itself. In this framework points of the plane are interpreted as being given by circles with radius $0$.

 Later 
the mathematician August Ferdinand M\"obius (1790-1868) called the joint sets of circles, straight lines and their mappings
by reciprocal radii ``Kreisverwandtschaften'' (circle relations) \cite{moeb}. Behind this notion is an implicit
 characterization of group theoretical properties which were only identified explicitly later when group theory for
 continuous transformation groups became established. 

Analytically the ``Kreisverwandtschaften'' are characterized by  the transformation formulae \eqref{eq:10} (nowadays called 
``M\"obius transformations''):
\begin{equation}
  \label{eq:38}
  z \to \hat{z} = \frac{a\,z + b}{c\,z + d}, \quad z = \frac{d\,\hat{z} - b}{- c\,\hat{z}
 + a},\quad a\,d- b\,c \neq 0,
\end{equation}
implying
\begin{equation}
  \label{eq:39}
  (d\,\hat{x})^2 + (d\,\hat{y})^2 = \frac{|a\, d- b\, c|^2}{|c\,z +d |^4}\,[(dx)^2 + (dy)^2].
\end{equation}
Multiplying numerators and denominators in Eq.\ \eqref{eq:38} by an appropriate complex  number one can normalize the
coefficients such that 
\begin{equation}
  \label{eq:43}
  a\, d - b\, c = 1.
\end{equation}
The last equation implies that 6 real parameters of the 4 complex numbers $a,\ldots,d$ are independent. In group theoretical
language: the transformations \eqref{eq:38} form a 6-dimensional group. 

Important special cases  are the linear transformations
\begin{equation}
  \label{eq:40}
  z \to \hat{z} =  a\,z + b,
\end{equation}
consisting of (2-dimensional) translations $T_2[b]\,$,  a (1-dimensional) rotation $D_1[\phi]$ 
 and a (1-dimensional) scale transformation (dilatation) $S_1[\gamma]\,$:
\begin{equation}
  \label{eq:224}
  T_2[b]:\, z \to z + b,\,b= b_1 + i\,b_2;\,\,D_1[\phi]: z \to e^{i\phi}z,\,\phi = \arg a; \,\, S_1[\gamma]: z \to
e^{\gamma}z,\,e^{\gamma} = |a|.
\end{equation}
 \\ Of special interest here is the discrete transformation
\begin{equation}
  \label{eq:41}
\bar{R}\,: \quad  z \to \hat{z} = \frac{r_0^2}{z} = \frac{r_0^2}{x^2 + y^2}\,(x-i\,y).
\end{equation}
This is the inversion \eqref{eq:31} followed by a reflection with respect to the $x$-axis. Notice that the r.h.\ side
$1/z$ is a {\em meromorphic} function on the complex plane with a pole at the point $z=0$ that is mapped onto the ``point''
$\infty$ which has to be ``joined'' to the complex plane. 
 
Another analytical implementation of the ``Kreisverwandtschaften'' is
\begin{equation}
  \label{eq:47}
   z \to \hat{z} = \frac{a\,z^* + b}{c\,z^* + d}, 
 \quad a\,d- b\,c =1 ,\quad z^* = x - i\,y,
\end{equation}
with an obvious corresponding expression for the relation \eqref{eq:39}.

The inversion \eqref{eq:31}
itself is given by
\begin{equation}
  \label{eq:42}
 R\,: \quad \hat{z} = \frac{r_0^2}{z^*},\quad z^* = x-i\,y.
\end{equation}
Here the orientation of angles is inverted, contrary to the transformation \er{eq:41}.

The combination $C_2[\beta] = R\cdot T_2[\beta]\cdot R $, where $T_2[\beta]$ denotes the translations $ z \to z+\beta,\,
\beta = \beta_1 + i\,\beta_2$, yields
\begin{equation}
  \label{eq:44}
 R\cdot T_2[\beta]\cdot R= C_2[\beta]\,: \quad z \to \hat{z} = \frac{z + \beta |z|^2}{1+2(\beta_1\,x +\beta_2\,y)
 +|\beta|^2\,|z|^2},
\end{equation}
which constitutes another 2-dimensional abelian subgroup, because $R^2 =1$ and $T_2[\beta]$ is abelian.

It is instructive to see which transformation is induced on the sphere of radius $a$  by the inverse 
stereographical projection \eqref{eq:24} when applied to the inversion \eqref{eq:42}. We can write the equations \er{eq:24}
 as
\begin{equation}
  \label{eq:45}
  \sigma = \xi + i\, \eta = \frac{4\,a^2\,z}{4\,a^2 +|z|^2},\quad \zeta = \frac{2a\,|z|^2}{4a^2 + |z|^2}.
\end{equation}
Taking for convenience $a= 1/2$ and $r_0 =1$, we get  for the inversion \eqref{eq:42}:
\begin{equation}
  \label{eq:46}
  \sigma \to \hat{\sigma} = \frac{\hat{z}}{1+|\hat{z}|^2} = \frac{z}{1+|z|^2} = \sigma, \quad \zeta \to \hat{\zeta} = 
\frac{|\hat{z}|^2}{1+|\hat{z}|^2} = \frac{1}{1+|z|^2} = 1-\zeta,
\end{equation}
i.e.\ the points $(\xi,\eta,\zeta)$ on the sphere are reflected on the plane $\zeta = 1/2$. The south pole of the
sphere which corresponds to the origin $(x=0,y=0)$ of the plane is mapped onto the north pole which corresponds to the
point $\infty$ of the plane. And vice versa. 

 Contrary to the non-linear transformation \eqref{eq:42} the transformation \eqref{eq:46} is a (inhomogeneous) linear
 and continuous one for the
coordinates of the sphere. We shall see below that such a linearization is possible for all the transformation
\eqref{eq:38} or \eqref{eq:47} by introducing appropriate homogeneous coordinates!

A few remarks on a more modern aspect: If $m(\tau),\,m(\tau = 0) = 1\,( \text{group identity}),$ denotes the elements of
 any of the above 6
real  one-parameter transformation subgroups and $f(z=x+i\,y)$ a smooth function on the complex plane, then
 \begin{equation} 
   \label{eq:48}
   \tilde{V}f(z) = \lim_{\tau \to 0}\frac{f[m(\tau)\,z]-f(z)}{\tau}
 \end{equation}
defines a vector field on the plane. In terms of the coordinates $x$ and $y$ these are for the individual groups
\begin{equation}
  \label{eq:49}
  T_2[b]\,:\quad \tilde{P}_x = \partial_x,\,\,\,\tilde{P}_y= \partial_y;
\end{equation}
\begin{equation}
  \label{eq:50}
  D_1[\phi]\,: \quad \tilde{L} = x\,\partial_y - y\,\partial_x;
\end{equation}
\begin{equation}
  \label{eq:51}
  S_1[\gamma]\,: \quad \tilde{S} = x\,\partial_x + y\,\partial_y;
\end{equation}
\begin{equation}
  \label{eq:52}
  C_2[\beta]\,: \quad \tilde{K}_x = (y^2-x^2)\,\partial_x -2x\,y\,\partial_y,\,\,\,\tilde{K}_y =(x^2-y^2)\,\partial_y
- 2x\,y\,\partial_x.
\end{equation}
These vector fields form a Lie algebra which is isomorphic to the real Lie algebra of the  M\"{o}bius group:
\begin{eqnarray}
  \label{eq:53}
 {[}\tilde{P}_x,\,\tilde{P}_y{]} &=& 0, \\
{[}\tilde{L},\,\tilde{P}_x{]}  & =& -\tilde{P}_y, \quad  {[}\tilde{L},\,\tilde{P}_y]  = \tilde{P}_x; \label{eq:54}
 \\
{[}\tilde{S},\,\tilde{P}_x{]} &=& - \tilde{P}_x,\quad {[}\tilde{S},\,\tilde{P}_y] = - \tilde{P}_y; \label{eq:55} \\
{[}\tilde{S},\,\tilde{L}{]} &=& 0\,; \label{eq:56} \\
{[}\tilde{L},\,\tilde{K}_x {]} &=& -\tilde{K}_y, \quad {[}\tilde{L},\,\tilde{K}_y {]} = \tilde{K}_x; \label{eq:61}\\
{[}\tilde{S},\, \tilde{K}_x{]} &=& \tilde{K}_x, \quad {[}\tilde{S},\, \tilde{K}_y] = \tilde{K}_y; \label{eq:57} \\
{[}\tilde{P}_x,\,\tilde{K}_x{]} &=& {[}\tilde{P}_y,\,\tilde{K}_y{]}= -2\,\tilde{S}; \label{eq:58} \\
{[}\tilde{P}_x,\,\tilde{K}_y{]} &=& -{[}\tilde{P}_y,\,\tilde{K}_x{]} = 2\,\tilde{L}. \label{eq:59}
\end{eqnarray}
Here we have already several of the essential structural elements of conformal groups we shall encounter later:
\begin{enumerate} \item The dilatation operator $\tilde{S}$ determines the dimensions of length of the operators
 $\tilde{P}_i,\,\tilde{L}\,$ and $\tilde{K}_i,i=x,y\,$, namely $[L^{-1}],\,[L^0]\,$ and $[L^1]$ as expressed by the
Eqs.\ \eqref{eq:55}, \eqref{eq:56} and \eqref{eq:57}. \item The Eqs.\ \eqref{eq:58} and \eqref{eq:59} show that the
 Lie algebra generators
$\tilde{L}$ and $\tilde{S}$ can be obtained from the commutators of $\tilde{P}_i$ and $\tilde{K}_i,\,i=x,y\,$. But
we know from Eq.\ \eqref{eq:44} that
\begin{equation}
  \label{eq:60}
  \tilde{K}_i = R\cdot \tilde{P}_i\cdot R,
\end{equation}
which means that the whole Lie algebra of the M\"{o}bius group can be generated from the translation generators 
$\tilde{P}_i$ and the inversion $R$ alone! \item 
We further have the relations
\begin{equation}
  \label{eq:62}
  R\cdot \tilde{S} \cdot R = - \tilde{S}, \quad  R\cdot \tilde{L} \cdot R =  \tilde{L}.
\end{equation} \end{enumerate}
 These properties indicate the powerful role of the discrete transformation $R$, mathematically
and physically! An appropriate name for $R$ would be ``length inversion (operator)''!

Many  properties of the inversion \eqref{eq:31} for the plane were investigated for the 3-dimensional
space, too, by the authors mentioned above (Durrande, Steiner, Pl\"{u}cker, ..., M\"{o}bius etc.), without realizing,
 however, that in
3 dimensions the transformation by reciprocal radii was essentially the only non-linear conformal mapping, contrary to
the complex plane and its extensions to Riemann surfaces with their wealth of holomorphic and meromorphic functions. 
This brings us to the next Subsect.\ :
\subsection{William Thomson, Joseph Liouville, Sophus Lie, \\ other  mathematicians and
James Clerk Maxwell}
Prodded by his ambitious father, the mathematics professor James Thomson (1786-1849), in January of 1845 the young
William Thomson (1824-1907) -- later Baron Kelvin of Largs -- spent four and a half months in Paris in order to get
 acquainted, study and work with the well-known mathematicians there \cite{thomp}. His best Paris contacts Thomson had with
Joseph Liouville (1809-1882) whose prot\'{e}g\'{e} he became \cite{luet}. Back in Cambridge, in October 1845 Thomson wrote
Liouville a letter in which he proposed to use the relation \eqref{eq:32} for a sphere of radius $r_0$ in oder to solve
certain (boundary) problems in electrostatics, refering to discussions the two had in Paris. Excerpts from that letter
were published immediately by Liouville in the journal he edited \cite{thom1}. In June and September 1846 Thomson sent
two more letters excerpts of which Liouville published in 1847 \cite{thom2}, directly followed by a long commentary 
by himself \cite{liou2}.  In the first of these letters Thomson introduced the mapping
\begin{equation}
  \label{eq:63}
R\,: \;\;  x\; \to \; \xi =  \frac{x}{x^2 +y^2 +z^2}, \;\;  y\; \to \; \eta =  \frac{y}{x^2 +y^2 +z^2}, \;\; 
 z\; \to \; \zeta =  \frac{z}{x^2 +y^2 +z^2},  
\end{equation}
and pointed out that the function $\hat{h}(\vec{x}) = h(\vec{x}/r)/r,\, r= (x^2 +y^2 +z^2)^{1/2}$ is a solution of
the Laplace equation \eqref{eq:5}, if $h(\vec{x})$ is a solution. In his commentary Liouville discussed in detail several
properties of the mapping \eqref{eq:63} and gave it the name ``transformation {\em par rayons vecteurs r\'{e}ciproques},
relativement \`a l'origine O'' (italics by Liouville), from which the usual expression ``transformation by reciprocal
 radii'' derives.

 Afterwards Liouville made the important discovery that the transformation \eqref{eq:63}, combined with translations,
 is actually
the only generic conformal transformation in $\mathbb{R}^3$, contrary to the situation in the plane \cite{mon2}!

Liouville's result kindled a lot of fascination  among mathematicians and brought quite a number of
generalizations and new proofs:

 At the end of a paper by Sophus Lie (1842-1899), presented by A. Clebsch in April 1871
to the Royal Society of Sciences at G\"{o}ttingen, Lie concluded that the orthogonal transformations and those by reciprocal
radii belong to the most general ones which leave the quadratic form
\begin{equation}
  \label{eq:64}
  \sum_{\nu = 1}^n (dx_{\nu})^2 = 0 
\end{equation}
invariant \cite{lie1}. He does not mention the condition $ n > 2$ nor does he quote Liouville's proof for $n =3$.
  In a long
paper from October and November 1871 Lie gave a different proof for Liouville's theorem (which he quotes now) and points
 out in a footnote
that the results of  his paper Ref.\ \cite{lie1} imply a corresponding generalization for arbitrary $n > 2$ \cite{lie2}. 
He provided the details of the proof for $ n >2$ in an article from 1886 \cite{lie3} and in volume III of his ``Theorie der 
Transformationsgruppen'' \cite{lie4} from 1893. In the mean-time other proofs for the general case $n> 2$ had appeared: one
by a German  secondary school teacher, R.\ Beez \cite{bee}, and another one by Gaston Darboux (1842-1917) \cite{dar1}. 
For a more modern one see Ref.\ \cite{lev}.

For the case $n=3$ there are about a dozen new proofs of Liouville's theorem till around 1900 
\cite{goup,maxw1,bian,cap,gour,tait,sche,giac,fors,camp,brom,dar2}.

Most important, however, for the influence of Thomson's work on the physics community was that James Clerk Maxwell
 (1831-1879) devoted a whole
chapter in his ``Treatise'' to applications of the inversion -- combined with the notion of virtual electric images -- in
 electrostatics \cite{maxw2}. Maxwell's high opinion of Thomson's work is also evident from his review (in Nature) 
\cite{maxw3} of the reprint volume of Thomson's papers \cite{ thom2}. Maxwell says there: \\ `` ... Thus Thomson
 obtained the
rigorous solution of electrical problems relating  to spheres by the introduction of an imaginary electrified system
within the sphere. But this imaginary system itself next became the subject of examination, as the result of the 
transformation of the external electrified system by reciprocal {\em radii vectores}. By this method, called that of
electrical inversion, the solution of many new problems was obtained by the transformation of problems already solved. ...
If, however, the mathematicians were slow in making use of the physical method of electric inversion, they were more
ready to appropriate the geometric idea of inversion by reciprocal {\em radii vectores}, which is now well known to all
geometers, having been, we suppose, discovered and re-discovered repeatedly, though, unless we are mistaken, most of these
discoveries are later than 1845, the date of Thomson's paper. ...'' \cite{neu}.
\subsection{Gaston Darboux and the linear action of the \\ conformal group on ``polyspherical'' 
coordinates}
We now come to a {\em global} aspect of the action of the conformal group which plays a major role in the modern development
of its applications (see Subsects.\ 7.2 and 7.4 below): We have already seen that the mapping \eqref{eq:31} sends the
 origin of the plane to infinity and
vice versa. Similar to what is being done in projective geometry where one adds an ``imaginary'' straight line at 
infinity one
now adds a point at infinity  in order to have the mapping \eqref{eq:31} one-to-one. Topologically this means that
 one makes
the non-compact plane to a compact 2-dimensional surface $S^2$ of the sphere. This is implemented by the stereographic
 projection
\eqref{eq:24}. As the projection is conformal it preserves an essential part of the Euclidean metric structure of the
plane, e.g.\ orthogonal curves on the sphere are mapped onto orthogonal curves in the plane. In addition the non-linear
 transformations \eqref{eq:42} become linear on  $S^2$ if one introduces homogeneous
coordinates in the plane and in the associated space $\mathbb{R}^3$ in which the sphere $S^2$ is embedded, i.e.\ 
the conformal transformations act continuously on $S^2$. This does not seem to be very exciting for the plane and the
sphere $S^2$, but it becomes important for the Minkowski space \er{eq:1} where the inversion \eqref{eq:3} is
 singular on
the 3-dimensional light cone $(x,x) =0$. But the essential ingredients of the idea can already be seen in the case of
the plane and the sphere $S^2$ which also shows the close relationship between stereographic projections and mappings
be reciprocal radii!
\subsubsection{Tetracyclic coordinates for the compactified plane}
In a short note from 1869 Darboux pointed out \cite{dar3} that one could generate a system of orthogonal curvilinear
 coordinates in the plane by projecting them stereographically from a given system on the surface of a sphere in space.
More generally, he observed that properties of an $\mathbb{R}^{n-1}$ could be dealt with by considering the corresponding
properties on the $(n-1)$-dimensional surface $S^{n-1}$ of a sphere in an $\mathbb{R}^n$. He discussed the details
 for $n = 3,4$
in later publications, especially in his monograph of 1873 \cite{dar4}. The following is a brief summary of the main
ideas, using also later textbooks on the subject \cite{poc1,boch,klei,enc}:

First one introduces homogeneous coordinates on $\mathbb{R}^2$ and $\mathbb{R}^3$:
\begin{eqnarray}
  \label{eq:73}
 && x=y^1/k,\, y= y^2/k, \quad (y^1,y^2,k) \neq (0,0,0); \\ && \xi = \eta^1/\kappa,\,\,\eta = \eta^2/\kappa ,\,
 \zeta = \eta^3/\kappa, \quad (\eta^1,\eta^2,\eta^3,\kappa) \neq (0,0,0,0). \nonumber
\end{eqnarray}
Mathematically the new homogeneous coordinate $k$ is just a real number which -- in addition -- can be given an obvious
 physical
interpretation \cite{ka1,ka2}: As the coordinates $x$ and $y$ have the dimension of length, one can interpret $k$ as
 providing
the length scale by giving it the dimension $[L^{-1}]$ so that the coordinates $y^1$ and $y^2$ are dimensionless.
For the sphere from Eq.\ \er{eq:23} we get now
\begin{eqnarray}
  \label{eq:74}
 Q(\vec{\eta},\,\kappa) &\equiv& (\eta^1)^2 + (\eta^2)^2 + (\eta^3)^2 - 2(a\,\kappa)\,\eta^3 = \\
 &=& 
 (\eta^1)^2 + (\eta^2)^2  -2 \eta^3\,\chi \equiv  Q(\vec{\eta}, \, \chi)   =   0, \nonumber \\ && \chi
 = a\,\kappa - \eta^3/2, \quad \vec{\eta} =  (\eta^1, \eta^2, \eta^3). \nonumber
\end{eqnarray}
Here a corresponding physical dimensional interpretation of $\kappa$ is slightly more complicated as the system has
already the intrinsic fixed length $a\,$: Now -- like $k$ in the plane -- the carrier of the  dimension  of an
 inverse length is the coordinate $\chi$ 
from Eqs.\ \er{eq:74}. This follows immediately from
the transformation formulae \er{eq:23} which may  be written as
\begin{equation}
  \label{eq:75}
  \sigma\, y^1 = \eta^1, \quad \sigma \, y^2 = \eta^2, \quad \sigma\, (a\,k) = \chi,
 \quad
\sigma \neq 0.
\end{equation}
Here $\sigma$ is an arbitrary non-vanishing real number which drops out when the ratios in Eqs.\ \er{eq:23} or \er{eq:73}
 are formed. It follows that
\begin{equation}
  \label{eq:76}
  \eta^3 = \sigma \,\, \frac{(y^1)^2 + (y^2)^2}{2a\,k}, \quad a\,\kappa = \sigma\,\left(a\,k +\frac{(y^1)^2 +
 (y^2)^2}{4a\,k}\right).
\end{equation}
We could also start from the Eqs.\ \er{eq:24} and get
\begin{eqnarray}
  \label{eq:77}
  \rho\,\xi &=& 4 (a\,k)\, y^1,\quad  \rho\,\eta = 4 (a\,k)\, y^2,\quad \rho\,\zeta = 2 [(y^1)^2 + (y^2)^2], \\
\rho\,(a\,\kappa)& =& 4 (a\,k)^2 + (y^1)^2 + (y^2)^2,\quad \rho \neq 0. \nonumber
\end{eqnarray}
The two formulations coincide for $\rho\,\sigma = 4(a\,k)$. 

We see that we can characterize the points in the plane -- including the ``point''  $\infty$ -- by 3 ratios of 4
 homogeneous coordinates which in addition obey the bilinear relation
\begin{equation}
  \label{eq:111}
   Q(\vec{\eta},\,\chi)=0.
\end{equation}
It follows from Eqs.\ \er{eq:73} that the point $\infty$ lies on the projective straight line $k=0$. According to 
Eqs.\ \er{eq:75} and \er{eq:74} this implies $\eta^3/\kappa = 2a$ and $(\eta^1,\eta^2)= (0,0)$, i.e.\ the
coordintes of the north pole.

The action of the different subgroups of the M\"{o}bius group as discussed in Subsect.\ 2.2 on the homogeneous
coordinates \er{eq:73} may be described as follows:

The {\em scale transformation}
\begin{equation}
  \label{eq:78}
 S_1{[}\gamma {]}\,: \quad z \to z^{\prime} = e^{\gamma}\, z
\end{equation}
can be implemented by 
\begin{equation}
  \label{eq:87}
\tau \, y^{1\,\prime} = y^1, \quad \tau\, y^{2\,\prime} = y^2, \quad \tau \, k^{\prime} = e^{-\gamma}\,k,
\end{equation}
where $\tau$ again is an arbitrary real number $ \neq 0$. 
 As $y^1$ and $y^2$ are dimensionless and $k$ has the dimension of an inverse length, properties which should not be
changed by the transformation, we put $\tau = 1\,$.
As to such a choice of the, in principle, arbitrary real number $\tau \neq 0$ see below.  
It then follows from Eqs.\ \er{eq:75}  that
\begin{equation}
  \label{eq:80}
\chi \equiv  a\,\kappa - \eta^3/2\, \to \, \chi^{\prime} = e^{-\gamma}\,\chi,
\end{equation}
i.e.\ the combination $\chi$ has the dimension of an inverse length. We have $Q(\vec{\eta}^{\,\prime},\,\chi^{\prime})
=Q(\vec{\eta},\,\chi)\,$.

For the {\em translation}
\begin{equation}
  \label{eq:81}
 T_2{[}b_1{]}\,: \quad x \to x^{\prime} = x +b_1, \quad y \to y^{\prime} =  y,
\end{equation}
we obtain accordingly
\begin{equation}
  \label{eq:82}
  \tau \, y^{1\,\prime} = y^1 +b_1\,k, \quad \tau\, y^{2\,\prime} = y^2, \quad \tau \, k^{\prime} = k.
\end{equation}
Again taking $\tau =1$ we get
\begin{eqnarray} \eta^1 \to \eta^{1\,\prime}& =& \eta^1 + (b_1/a)\,\chi, \quad
\eta^2 \to \eta^{2\,\prime} = \eta^2, \label{eq:84} \\
\eta^3 \to \eta^{3\,\prime} &=& (b_1/a)\,\eta^1 + \eta^3 + (b_1/a)^2/2\,\chi, \quad
\chi \to \chi^{\prime} = \chi. 
\end{eqnarray}
This transformation also leaves the quadratic form $Q(\vec{\eta},\, \chi)$ invariant: $Q(\vec{\eta}^{\,\prime},\,
 \chi^{\prime})
=Q(\vec{\eta},\, \chi)\,$. \\
The translation $ x \to x,\, y \to y + b_2$ can be treated in the same way. 

For the {\em inversion}
\begin{equation}
  \label{eq:88}
 R\,: \quad x \to x^{\prime} = \frac{r_0^2\,x}{x^2 + y^2}, \quad  y \to y^{\prime} = \frac{r_0^2\,y}{x^2 + y^2},
\end{equation}
one obtains
\begin{equation}
  \label{eq:89}
  y^{1\,\prime} = y^1, \quad y^{2\,\prime} = y^2, \quad k^{\prime} = \frac{(y^1)^2 + (y^2)^2}{
r_0^2 \,k}.
\end{equation}
This yields
\begin{equation}
  \label{eq:90}
  R\,: \quad \eta^{1\,\prime} = \eta^1, \quad \eta^{2\,\prime} = \eta^2, \quad  
y^{3\,\prime} = \frac{r_0^2}{2\, a^2}\, \chi, \quad \chi^{\prime} = \frac{2\,a^2}{r_0^2}\,\eta^3.  
\end{equation}
We again have $Q(\vec{\eta}^{\,\prime},\, \chi^{\prime})=Q(\vec{\eta},\, \chi)$.
Invariance of $Q$ under rotations in the $(x,y)$--plane and the corresponding $(\xi,\eta)$--plane, with $k,\,\kappa$ and
$\zeta$ fixed, is obvious.

Introducing the coordinates
\begin{equation}
  \label{eq:94}
  \xi^1 = \eta^1,\quad \xi^2 = \eta^2,\quad \xi^3= \frac{1}{\sqrt{2}}(\chi + \eta^3),\quad
\xi^0= \frac{1}{\sqrt{2}}(\chi - \eta^3),
\end{equation}
implies
\begin{equation}
  \label{eq:95}
  Q(\vec{\eta},\,\chi) = Q(\xi,\xi) = (\xi^1)^2 +(\xi^2)^2 + (\xi^3)^2 - (\xi^0)^2.
\end{equation}
Thus, we see that the 6-dimensional conformal group of the plane -- M\"{o}bius group \er{eq:38}  with the normalization
\er{eq:43} -- is isomorphic to the 6-dimensional pseudo-orthogonal (``Lorentz'') group $O(1,3)/\mathbb{Z}_2$ (division by
$\mathbb{Z}_2\,: \xi \to \xi \text{ or } -\xi$, because the coordinates $\xi$ are homogeneous ones).
 The inversion $R$,
Eq.\ \er{eq:88}, e.g.\  is implemented by the ``time reversal'' $ \xi^0 \to -\xi^0$! As 
 $\xi$ is equivalent to $-\xi$,  ``time reversal'' here is equivalent
 to ``space
reflection'':  $\xi^0 \to \xi^0,\,\xi^j \to -\xi^j,\,j=1,\,2,\,3\,$.

The homogeneous coordinates $\vec{\eta},\,\kappa$  of Eq.\ \er{eq:73} or any linear combination of them,
 e.g.\ \eqref{eq:94},
were called ``{\em tetracyclic}'' coordinates of the points in the plane (the point $\infty$ included), i.e.\
 ``four-circle'' coordinates (from the Greek words
``tetra'' for four and ``kyklos'' for circle) \cite{klei,enc,poc1,boch}. The geometrical background for this name is
 the following: We have seen above,
 Eq.\ \er{eq:65}, that a circle on the sphere may be characterized by the plane passing through the circle. In homogeneous
coordinates  Eq.\ \er{eq:65} becomes
\begin{equation}
  \label{eq:96}
  c_1\,\eta^1 +c_2\, \eta^2 + c_3\,\eta^3 +c_0\,\kappa = 0.
\end{equation}
The four different  planes $\eta^1 = 0\,;\, \eta^2=0\,;\,\eta^3=0$ or $\kappa = 0$ correspond to four circles in
the plane (which may have radius $\infty\,$, i.e.\ they are straight lines). On the other hand, let, in the notation
 of Eq.\ \er{eq:33},
\begin{equation}
  \label{eq:97}
  B_j \equiv  (x-\alpha_j)^2 + (y-\beta_j)^2 - \rho_j^2 = x^2 + y^2 -2\alpha_j \,x - 2\beta_j \, y +C_j = 0,\,
\, j=1,2,3,4,
\end{equation}
be four different and arbitrary circles in the plane, each of which is determined by three parameters $\alpha_j,\,\beta_j$
and $C_j$ or the radius $\rho_j$. Any other circle $B=0$ in the plane can be characterized by the relation
\begin{equation}
  \label{eq:98}
  B= \sum_{j=1}^4 \eta^j\,B_j = 0, \quad (\eta^1,\eta^2,\eta^3,\eta^4) \neq (0,0,0,0), 
\end{equation}
which constitute 4 homogeneous equations for the 3 inhomogeneous parameters of the fifth circle. Its radius squared
$\rho^2$ becomes proportional to a bilinear form of the homogeneous coordinates $\eta^j$.
 If the new circle \er{eq:98} is a point, i.e.\ $\rho=0\,$, then
the $\eta^j$ obey a quadratic relation like \er{eq:74} or \er{eq:95} (with $Q(\xi,\xi)= 0\,$). This is the geometrical
 background
for the term ``tetracyclic'' coordinates for points in the plane. It is a variant of the term ``pentaspherical'' coordinates
originally introduced by Darboux in the corresponding case of characterizing points in 3-dimensional space in terms of five
(Greek: ``penta'') homogeneous coordinates which obey a bilinear relation \cite{dar5}. 
\subsubsection{Polyspherical coordinates for the extended $\mathbb{R}^n,\,n \geq 3$}
Let $x^{\mu},\,\mu =1,\ldots,n$ be the cartesian coordinates of an $\mathbb{R}^n$ with the bilinear form \er{eq:22}.
Then, without refering to an explicit $(n+1)$-dimenional geometrical background, so-called ``polyspherical'' coordinates
 $y^{\mu},\, \mu =
1,\ldots ,n,\,k$ and $q\, $, can be introduced
by
\begin{equation}
  \label{eq:99}
  x^{\mu} = y^{\mu}/k ,\quad (y,y) - k\,q = 0.
\end{equation}
Here $k$ has the dimension of an inverse length, and $q$ that of a length. 

The conformal transformations in such a $\mathbb{R}^n$ consists of $n$ translations $T_n[b]$, $n\,(n-1)/2$
 pseudo-rotations $D_{n(n-1)/2}[\phi_{\nu}]$, one scale transformation $S_1[\gamma]$, $n$ ``special conformal'' 
transformations of the type
\er{eq:44}: $\,C_n[\beta] = R\cdot T_n[\beta]\cdot R\,$ and discrete transformations like $R$ etc.
 Combined these make a transformation group of
 dimension $(n+1)(n+2)/2$. 

If the bilinear form \er{eq:22}  is a ``lorentzian'' one,
\begin{equation}
  \label{eq:79}
  (x,x) = (x^0)^2 - (x^1)^2 - \cdots - (x^{n-1})^2,
\end{equation}
then the associated bilinear form \er{eq:99} is
\begin{equation}
  \label{eq:100}
  Q(y,y) = (y^0)^2+(y^{n+1})^2 - (y^1)^2 - \cdots - (y^n)^2, \quad k = y^n + y^{n+1},\,\, q= y^n - y^{n+1}.
\end{equation}
In this case one would properly speak of ``poly-hyperboloidical'', and for $n=4$ of ``hexa-hyperboloidical'' 
coordinates (``hexa'': Greek for six)!

The conformal group of the $n$-dimensional Minkowski space now coresponds to the group $O(2,n)/\mathbb{Z}_2\,$. 
Its global structure
and that of the manifold $Q(y,y) = 0$ will be discussed in Subsect.\ 7.2 below. As already discussed for the plane, the
 division by $\mathbb{Z}_2$ comes from
 the fact that
one can multiply the homogeneous coordinates $y^{\nu}$ in  Eq.\ \er{eq:100} by an arbitrary real number $\rho \neq 0$ 
without affecting the coordinates $x^{\mu}$ in Eq.\ \er{eq:99}.

If one now wants to discuss conformally invariant or covariant differential equations (``field equations'') of functions
$F(y)$ on the
 manifold $Q(y,y) =0\,$ one has to take into account the homogeneity of the coordinates $y$ in Eq.\ \er{eq:99} and
the condition $ (y,y) -k\,q = 0$. The work on this task was started by Darboux in the case of potential theory
 \cite{dar6}, extended by Pockels and B\^{o}cher \cite{poc1,boch}, later discussed by Paul Adrien Maurice Dirac (1902-1984)
\cite{dir1} and more recently by other authors, e.g.\ \cite{ka2,mac1,go2,rev1}. As the subject is more technical I refer to 
those papers and reviews for details.
\section{Einstein, Weyl and the origin of gauge theories}
\subsection{Mathematical beauty versus physical reality \\ and  the far-reaching consequences}
 In November 1915 Einstein had presented the final
 version of his relativistic theory of
gravitation in the mathematical framework of Riemannian geometry. Here the basic geometrical field quantities are the
 coefficients
$g_{\mu\nu}(x),\, \mu,\,\nu = 0,1,2,3 \,$ of the metric form (summation convention)
 
  \begin{equation}
    \label{eq:101}
     (ds)^2 = g_{\mu \nu}(x)\,dx^{\mu} \otimes dx^{\nu}.
  \end{equation}
The local lenghts $ds$ are assumed to be determined by physical measuring rods and clocks (made of atoms, molecules etc.).
A basic assumption of Riemannian geometry applied to gravity is that the physical units defined by those instruments 
are locally the same everywhere and at all  time, independent of the
gravitational fields present: we assume that hydrogen atoms etc.\ and their energy levels  locally do not differ from
each other everywhere in our cosmos and have not changed during its history. 

In 1918 Hermann Weyl proposed to go beyond this assumption in order to incorporate electromagnetism and its charge
conservation (for a more elaborate account of the following see the recent reviews \cite{orai1,orai2,schu,goe}):
 in Riemannian geometry {\em parallel transport} of a vector (``yardstick'') $ a = a^{\mu}\partial_{\mu}$
 does {\em not change its length} when brought from the point $P(x)$ to a neighbouring point $P(x+\delta x)$. This means
that $\delta [g_{\mu \nu}(x)\,a^{\mu} a^{\nu}] = 0$ (the covariant derivative of $g_{\mu \nu}$ vanishes, here formally
 characterized by $\delta g_{\mu \nu} = 0)$. 

Weyl now allows for a geometrical structure in which infinitesimal {\em parallel transport} of a vector can result in a
 change of length, too,  which is characterized by the postulate that this change is  given by
\begin{equation}
  \label{eq:102}
  \delta g_{\mu \nu}(x) = A(x)\, g_{\mu \nu}(x), \,\, A(x) = A_{\mu}(x)\,\delta x^{\mu}.
\end{equation}
For the Christoffel symbols of the first kind (which determine the parallel transport) this leads to the modification
\begin{eqnarray}
  \label{eq:103}
  \Gamma_{\lambda, \mu \nu} + \Gamma_{\mu, \lambda \nu} &=& \partial_{\nu}g_{\lambda \mu} + g_{\lambda \mu}\,A_{\nu}; \\
\Gamma_{\lambda, \mu \nu}&=& \frac{1}{2}(\partial_{\mu}g_{\lambda \nu} +\partial_{\nu}g_{\lambda \mu}-
\partial_{\lambda}g_{\mu \nu}) + \frac{1}{2}(g_{\lambda \mu}\,A_{\nu} +g_{\lambda \nu}\,A_{\mu}-g_{\mu \nu}\,A_{\lambda}).
\nonumber
\end{eqnarray}
The relation \er{eq:102} may be rephrased as follows: Let $l$ be the ``physical'' length $l=ds$ (Eq.\ \er{eq:101}) of 
a vector $a= a^{\mu}\partial_{\mu}$ at $P(x)\,$. If $a$ is parallel transported to a neighbour point 
 $P(x+ \delta x)$ then the change of its length $l$ is given
by
\begin{equation}
  \label{eq:105}
  \delta l = l\,A,
\end{equation}
which vanishes in Riemannian geometry. If one parallel transports a vector of  length $l_{P_1}$ from $P_1$ along a curve to 
$P_2$, then integration of Eq.\ \er{eq:105} gives the associated change
\begin{equation}
  \label{eq:106}
  l_{P_2} = l_{P_1}\,e^{\int_{P_1}^{P_2}A}.
\end{equation}
Here the integral is path-dependent if not all $F_{\mu \nu} = \partial_{\mu}A_{\nu} - \partial_{\nu}A_{\mu}$ vanish (Stokes'
theorem). 

Multiplying the metrical coefficients $g_{\mu \nu}(x)$ by a scale factor $\omega(x) > 0$ leads to the joint transformations:
\begin{equation}
  \label{eq:104}
 g_{\mu \nu}(x) \to \omega (x)\,g_{\mu \nu}(x), \quad A \to A - \delta \omega / \omega = A - \delta (\ln \omega),\,\,
\delta \omega = \partial_{\mu}\omega(x)\,\delta x^{\mu}.
\end{equation}

All this strongly suggests to identify (up to a constant) the four $A_{\mu}(x)$ with the electromagnetic potentials 
 and the transformation
\er{eq:104} with a gauge transformation affecting simultaneously both, gravity and electromagnetism. Making a special 
choice for $\omega$ Weyl called ``Eichung'' (= ``gauge''). In this way the term entered the realm of physics \cite{weyl1}.

 On March 1, 1918, Weyl wrote a letter to Einstein announcing a forthcoming paper on the unification of gravity and
electromagnetism and asking whether the paper could be presented by Einstein to the Berlin Academy of Sciences
 \cite{weyl2}. Einstein reacted enthusiastically on March 8 and promised to present Weyl's paper  \cite{weyl3}.
After receiving it he called it (on April 6) ``a first rank stroke of a genius'', but that he could not get rid of his 
``Massstab-Einwand'' (measuring rod objection) \cite{weyl4}, probably alluding to discussions the two had at the end of
 March in Berlin during Weyl's visit. 

 This was the beginning of a classical controversy over mathematical beauty
versus physical reality! On April 8 Einstein wrote ``apart from its agreement with reality it is in any case a superb
achievement of thought'' \cite{weyl5}, and again on April 15: ``As beautiful as your idea is, I have to admit openly that
according to my view it is impossible that the theory corresponds to nature'' \cite{weyl6}. Einstein's first main objection
concerned the relation \er{eq:106}: In the presence of electromagnetic fields $F_{\mu \nu}$ two ``identical'' clocks
could run differently after one of them was moved around on a closed path! Einstein communicated his physical objections
in a brief appendix to Weyl's initial paper he presented to the Academy in May 1918 \cite{weyl7}.
 The lively exchange between 
Einstein and Weyl continued till the end of the year, with Weyl trying hard to persuade Einstein. To no avail: on Sept.\
27 Einstein wrote: ``How I think with regards to reality you know already; nothing has changed that. I know how much 
easier it is to persuade people, than to find the truth, especially for someone, who is such an unbelievable master of
depiction like you.'' \cite{weyl8}. In a letter from Dec.\ 10 Weyl said, disappointed: `` So I am hemmed in between the
belief in your authority and my insight. ... I simply cannot otherwise, if I am not to walk all over my mathematical
conscience'' \cite{weyl9}. Einstein's answer from Dec.\ 16 is quite conciliatory: ``I can only tell you that all I 
talked to, from a mathematical point of view spoke with the highest admiration about your theory and that I, too, admire
it as an edifice of thoughts. You don't have to fight, the least against me. There can be no question of anger on my side:
Genuine admiration but unbelief, that is my feeling towards the matter.'' \cite{weyl10}. Einstein's other main objection
 concerned the relation \er{eq:103} which determines geodetic motions: it implies that an uncharged particle would
 nevertheless be influenced by an electromagnetic field!

Einstein was right as far as gravity and classical electrodynamics is concerned. But Weyl's idea found an unexpected
rebirth and modification in the quantum theory of matter \cite{orai1,orai2,schu,goe}: In 1922 Erwin Schr\"{o}dinger
 (1887-1961) observed -- by
discussing several examples -- that the Bohr-Sommerfeld quantization conditions are compatible with Weyl's gauge factor
\er{eq:106} if one replaces the  real exponent by the imaginary one
\begin{equation}
  \label{eq:107}
  \frac{i\,e}{\hbar}\,\int A,\qquad A = A_{\mu}dx^{\mu},
\end{equation}
where the $A_{\mu}(x)$ are now the usual electromagnetic potentials \cite{schr}. In 1927 Fritz London (1900-1954) 
reinterpreted Weyl's theory in the framework of the new wave mechanics \cite{lon}: like Schr\"{o}dinger, whom he quotes,
 London replaces the real exponent in Weyl's gauge factor \er{eq:106} by the expression \er{eq:107} and assumes
 that a length
$l_0$ when transported along a closed curve in a nonvanishing electromagnetic field acquires a phase change
\begin{equation}
  \label{eq:108}
  l_0 \to l=l_0\,e^{(ie/\hbar)\int A},
\end{equation}
without saying why a length could become complex now. He then argues -- in a way which is difficult to follow -- that
\begin{equation}
  \label{eq:109}
  \psi(x)/l(x) = |\psi|/l_0 = \text{const.},
\end{equation}
where $\psi(x)$ is a wave function which now posesses the phase factor from Eq.\ \er{eq:108}. 

 In two impressive papers from 1929 \cite{weyl11} and 1931 \cite{weyl12} Weyl himself revoked his approach
 from 1918 and reinterpreted his gauge transformations in the new quantum mechanical framework as implemented by
 \begin{equation}
   \label{eq:110}
   \psi \to e^{i\,e\,f/\hbar}\, \psi, \quad A_{\mu} \to A_{\mu} + \partial_{\mu}f,\quad \partial_{\mu} \to 
\partial_{\mu} - \frac{i\,e}{\hbar}\,A_{\mu},\quad f(x)= - \ln \omega(x),
 \end{equation}
giving credit to Schr\"{o}dinger and London in the second paper \cite{weyl12}. Especially this second paper with its
conceptually brilliant and broad analysis as to the importance of geometrical ideas in physics and mathematics provided
 the basis for the great future of gauge theories in physics and that of fiber bundles in mathematics.

Also stimulated by Weyl's idea, another interesting attempt to unify gravitation and electromagnetism was that of 
Theodor Kaluza (1885-1954) who started
from a 5--dimensional Einsteinian gravity theory with a compactified 5th dimension \cite{kal}. As to further developments
(O.\ Klein and others) of this approach see Ref.\ \cite{orai1}.
\subsection{Conformal geometries}
Despite its (preliminary) dead end in physics, Weyl's ideas were of considerable interest in mathematical differential
geometry. Weyl discussed them in several articles \cite{wey1} and especially, of course, in his textbooks \cite{wey2}.
 An important
new notion in these geometries was that of the (conformal) {\em weight} $e$ of geometrical quantities $Q(x,\, g_{\mu \nu})$
 like tensors or
tensor densities which depend on the $g_{\mu \nu}$ and their derivatives \cite{wey3}: The covariant metric tensor
 has the weight $1$, thus
\begin{equation}
  \label{eq:112} \text{if }\,\,
  g_{\mu \nu}(x) \to \omega(x)\, g_{\mu \nu}(x),\,\,\, \text{then } \quad Q(x,\, \omega\, g_{\mu \nu}) =
 \omega^e(x)\, Q(x,\, g_{\mu \nu}).
\end{equation}
Only quantities of weight $e=0$ are conformal invariants. \\
Even Einstein contributed to this kind of geometry \cite{ein1}, defined by the invariance of the bilinear
form
\begin{equation}
  \label{eq:113}
  g_{\mu \nu}(x)\,dx^{\mu} \otimes dx^{\nu} = 0,
\end{equation}
which also characterizes light rays and their associated causal cones. Einstein was interested in the relationship between
 ``Riemann-tensors'' and ``Weyl-tensors''. Most of the mathematical developments of these conformal geometries are
 summarized
in the second edition of a textbook by Jan Arnoldus Schouten (1883-1971) who himself made substantial contributions 
to the subject \cite{scho}. 

For a concise summary of conformal transformations in the sense of Weyl and their role in a modern geometrical
 framework of General Relativity
and associated field equations see, e.g.\ Ref.\ \cite{wald}.
\subsection{Conformal infinities}
We have seen in Subsect.\ 2.2 that it can have advantages to map the ``point'' $\infty$ and its neighbourhoods into a
finite one -- either by a  reciprocal radii transformation in the plane or by a stereographic projection onto the
 sphere $S^2$ --, if one wants to
investigate properties of geometrical quantities near $\infty$. Similarly, Weyl's conformal transformations \er{eq:112}
have been used to develop a sophisticated analysis of the asymptotic behaviour of space-time manifolds, especially for
those which are asymptotically flat \cite{coninf}, also possibly with a change of topological properties.
 They even have become an important tool for the {\em numerical} analysis
of black holes physics  etc.\ \cite{frau}.

Furthermore, as the rays of electromagnetic and gravitational radiation obey the relation \er{eq:113} and as these
rays form the boundaries (``light cones'') between causally connected and causally disconnected regions, Weyl's conformal
transformations play also an important role in the causal analysis of space-time structures \cite{caus}.
\section{Emmy Noether, Erich Bessel-Hagen and \\ the (partial) conservation of conformal currents}
\subsection{Bessel-Hagen's paper from 1921 on the conformal currents in electrodynamics}
I indicated already in Subsect.\ 1.1 that the form invariance of Maxwell's equations with respect to conformal space-time
transformations as discovered by Bateman and Cunningham does not necessarily imply new conservation laws. This point was
clarified in 1921 in an important paper by Erich Bessel-Hagen (1898-1946):

In July 1918 Felix Klein (1849-1925) had presented Emmy Noether's (1882-1935) seminal paper with her now two famous theorems
on the consequences of the invariance of an action integral either under an $r$-dimensional continuous (Lie) group or
under an ``infinite''-dimensional (gauge) group  the elements of which depend on  $r$  arbitrary functions
\cite{noe,ka3}. The former leads to $r$ conservation laws (first theorem),
 whereas the latter entails $r$ identities among the Euler-Lagrange expressions for the field equations
 (second theorem, e.g.\ the 4 Bianchi
 identities as a consequence of the 4 coordinate diffeomorphisms). 

In the winter of 1920 Klein encouraged  Bessel-Hagen
 to apply the first theorem to the conformal invariance of Maxwell's equations as discovered by Bateman and
Cunningham.

Bessel-Hagen's paper \cite{bes} contains a number of results which are still generic examples for modern applications of the
theorems: He first generalized Noether's results by not requiring the invariance of $L\,dx^1\cdots dx^m$ inside the action
integral, but by allowing for an additional total divergence $\partial_{\mu}b^{\mu}$ which is also linear in the
 infinitesimal group parameters
or (gauge) functions.
Because of the importance of Noether's first theorem let me briefly summarize its content: \\
Suppose the differential equations for $n$ fields $\vp^i(x),\,i=1,\ldots,n,\, x=(x^1,\ldots,x^m),$ are obtained from an
action integral
\begin{equation}
  \label{eq:114}
  A= \int_{G}dx^1\cdots dx^m\,L(x; \vp^i,\,\partial_{\mu}\vp^i).
\end{equation}
Let
\begin{equation}
  \label{eq:115}
  x^{\mu} \to \hat{x}^{\mu} = x^{\mu} +\delta x^{\mu}\,;\,\, \vp^i(x) \to \hat{\vp}^i(\hat{x}) = \vp^i(x) +
 \delta \vp^i = \vp^i(x) + \tilde{\delta}\vp^i + \partial_{\mu}\vp^i\,\delta x^{\mu},
\end{equation}
be infinitesimal transformations which imply
\begin{eqnarray}
  \label{eq:116}
  \delta A &=& \int_{\hat{G}}d\hat{x}^1\cdots d\hat{x}^m\,L[\hat{x}; \hat{\vp}^i(\hat{x}),\hat{\partial}
 \hat{\vp^i}(\hat{x})]-\int_{G}dx^1\cdots dx^m\,L[x; \vp^i(x),\partial\vp^i(x)] \\ \nonumber
&=& \int_{G}dx^1\cdots dx^m\,[E_i(\vp)\,\tilde{\delta}\vp^i -\partial_{\mu}j^{\mu}(x;\,  \vp,\,\partial\vp; \delta x, \, 
\delta \vp)], \end{eqnarray} \begin{eqnarray} \label{eq:127}
 E_i(\vp) &\equiv& \frac{\partial L}{\partial \vp^i} - \partial_{\mu}\frac{\partial L}{\partial
(\partial_{\mu}\vp^i)},  \\ \label{eq:128}
j^{\mu} &=& T^{\mu}_{\;\;\nu}\,\delta x^{\nu} - \frac{\partial L}{\partial \vp^i}\,\delta \vp^i +b^{\mu}(x; 
 \vp,\partial\vp; \delta x,  \delta \vp), \\ \nonumber
T^{\mu}_{\;\;\nu}&=& \frac{\partial L}{\partial(\partial_{\mu} \vp^i)}\,\partial_{\nu}\vp^i-\delta^{\mu}_{\,\,\nu}\,L. 
\end{eqnarray}
From $\delta A = 0$ and since the region $G$ is arbitrary we get the general variational identity
\begin{equation}
  \label{eq:126}
  \partial_{\mu}j^{\mu} = -E_i(\vp)\,\tilde{\delta}\vp^i.
\end{equation}
If 
\begin{equation}
  \label{eq:117}
  \delta x^{\mu} = X^{\mu}_{\,\rho}(x)\,a^{\rho}, \quad \delta \vp^i = \Phi^i_{\,\rho}(x,\vp)\,a^{\rho}, \quad
b^{\mu} = B^{\mu}_{\,\rho}(x,\vp,\partial \vp)\,a^{\rho}, \,\, |a^{\rho}| \ll 1,\,\, \rho =1, \ldots, r,
\end{equation}
where the $a^{\rho}$ are $r$ independent infinitesimal group parameters, then one has $r$ conserved currents
\begin{equation}
  \label{eq:118}
  j^{\mu}_{\,\rho}(x) = T^{\mu}_{\;\;\nu}\,X^{\nu}_{\,\rho} - \frac{\partial L}{\partial(\partial_{\mu}\vp^i)}\,\Phi^i_{\,\rho} +
B^{\mu}_{\,\rho},\,\,\rho = 1,\ldots,r,
\end{equation}
i.e.\ we have
\begin{equation}
  \label{eq:119}
  \partial_{\mu}j^{\mu}_{\rho}(x) = 0,\,\, \rho = 1,\ldots,r,
\end{equation}
 for solutions $\vp^i(x)$ of the field equations $E_i(\vp) = 0,\,i= 1,\ldots,n$.

As a first application Bessel-Hagen shows that such an additional term $b$ occurs for the $n$-body system in classical
 mechanics if one wants to derive the uniform center of mass motion from the 3-dimensional special Galilean group
$\vec{x}_j \to \vec{x}_j + \vec{u}\,t,\,j=1,\,\ldots,\,n\,$.

In electrodynamics (4 space-time dimensions)  he starts from the free Lagrangean density 
\begin{equation}
  \label{eq:120}
  L = -\frac{1}{4}\,F_{\mu \nu}F^{\mu \nu}, \quad F_{\mu \nu} = \partial_{\mu}A_{\nu} - \partial_{\nu}A_{\mu},\quad \frac{
\partial L}{\partial(\partial_{\mu}A_{\nu})} = - F^{\mu \nu}, \quad E^{\nu}(A)= \partial_{\mu}F^{\mu \nu}.
\end{equation}
Requiring the 1-form $A_{\mu}(x)dx^{\mu}$ to be invariant implies
\begin{equation}
  \label{eq:122}
  \delta A_{\mu} = - \partial_{\mu}(\delta x^{\nu})\, A_{\nu}(x).
\end{equation}
Notice that only those $\delta x^{\mu}$ and $\delta A_{\mu}$ are of interest here which leave  $L\,dx^0dx^1 
dx^2dx^3$ invariant (up to a total divergence) with $L$ from Eq.\ \er{eq:120}.
For the (infinitesimal) gauge transformations
\begin{equation}
  \label{eq:123}
  A_{\mu}(x) \to A_{\mu}(x) + \partial_{\mu}f(x),\quad \delta A_{\mu} = \partial_{\mu}f(x),
 \quad |f(x)| \ll 1, \quad f(x) = \ln \omega(x),
\end{equation}
Noether's second theorem gives the identity $\partial_{\nu}E^{\nu}(A)= \partial_{\nu}\partial_{\mu} F^{\mu \nu} =0\,$
(which implies charge conservation if $E^{\nu}(A) = j^{\nu}(x)$).
The ``canonical'' energy-momentum tensor
\begin{equation}
  \label{eq:124}
  T^{\mu}_{\;\; \nu} = - F^{\mu \kappa}\partial_{\nu}A_{\kappa} +\frac{1}{4}\,\delta^{\mu}_{\,\nu} F^{\kappa \lambda}
F_{\kappa \lambda}
\end{equation}
is not symmetric and not gauge invariant. Its relation to the symmetrical and gauge invariant energy-momentum
tensor $\Theta^{\mu \nu}$ is given by
\begin{equation}
  \label{eq:125}
  T^{\mu \nu} = \Theta^{\mu \nu} - F^{\mu \kappa}\partial_{\kappa}A^{\nu}, \quad \Theta^{\mu \nu} = F^{\mu \kappa}
F_{\kappa}^{\,\,\nu} + \frac{1}{4}\,\eta^{\mu \nu} F^{\kappa \lambda}F_{\kappa \lambda},
\end{equation}
where $\eta_{\mu \nu}$ represents the Lorentz metric from Eq.\ \er{eq:1}. \\
Combining the variations \er{eq:122} and \er{eq:123} with the relation \er{eq:125}
here gives for the current \er{eq:128} ($b^{\mu}=0$)
\begin{equation}
  \label{eq:130}
  j^{\mu} = \Theta^{\mu}_{\;\;\nu}\delta x^{\nu} + F^{\mu \nu}\partial_{\nu}(f-A_{\lambda}\delta x^{\lambda}).
\end{equation}
As $f(x)$ is an arbitrary function, Bessel-Hagen argues, we can choose the gauge
\begin{equation}
  \label{eq:131}
  f(x) = A_{\nu}\,\delta x^{\nu}
\end{equation}
for any given $\delta x^{\nu}$, so that now
\begin{equation}
  \label{eq:132}
  j^{\mu}= \Theta^{\mu}_{\;\;\nu}\delta x^{\nu},
\end{equation}
which is  invariant under another transformation \er{eq:123}.
For a gauge \er{eq:131} the variation $\tilde{\delta}A_{\mu}$ (see Eq.\ \er{eq:115}) takes the form
\begin{equation}
  \label{eq:133}
  \tilde{\delta}A_{\mu} = F_{\mu \nu}\delta x^{\nu},
\end{equation}
so that
finally
\begin{equation}
  \label{eq:134}
  \partial_{\mu}(\Theta^{\mu}_{\;\;\nu}\delta x^{\nu}) =  E^{\mu}(A)F_{\mu \nu}\delta x^{\nu}.
\end{equation}
For non-vanishing charged currents $j^{\mu}$ one has
\begin{equation}
  \label{eq:135}
  E^{\nu}(A)= \partial_{\mu}F^{\mu \nu} = j^{\nu}, \quad j^{\mu}F_{\mu \nu} = -F_{\nu \mu}j^{\mu} = - f_{\nu},
\end{equation}
where $f_{\nu}$ is the (covariant) relativistic force density. Eq.\ \er{eq:134} can therefore also be written as
\begin{equation}
  \label{eq:136}
   \partial_{\mu}(\Theta^{\mu}_{\;\;\nu}\delta x^{\nu}) = -f_{\nu}\delta x^{\nu}.
\end{equation}
Now let $S^{\mu \nu}$ be a symmetrical {\em mechanical} energy-momentum tensor such that
\begin{equation}
  \label{eq:137}
  \partial_{\mu}S^{\mu \nu} = f^{\nu},\quad S^{\mu \nu} = S^{\nu \mu},
\end{equation}
then Eq.\ \er{eq:136} may be rewritten as
\begin{equation}
  \label{eq:138}
  \partial_{\mu}[(\Theta^{\mu}_{\;\;\nu} + S^{\mu}_{\;\;\nu})\delta x^{\nu}] = S^{\mu}_{\;\;\nu}\partial_{\mu}(\delta x^{\nu}).
\end{equation}
This is an important equation from Bessel-Hagen's paper. 

As $\delta x^{\nu} = \text{ const.\ } $ for tranlations and
$\delta x^{\mu} = \omega^{\mu}_{\;\;\nu}x^{\nu},\,\omega_{\nu \mu} = -\omega_{\mu \nu},$ for homogeneous
Lorentz transformations, one sees immediately that the associated currents are conserved for the full system.
The situation is more complicated for the currents associated with the scale transformations
and the 4 special conformal transformations corresponding to those in Eq.\ \er{eq:44}:
\begin{eqnarray}
  \label{eq:121}
  S_1[\gamma]: x^{\mu} \to \hat{x}^{\mu} &=& e^{\gamma}\,x^{\mu},\quad \mu = 0,1,2,3, \\
 \delta x^{\mu} &=& \gamma\, x^{\mu}, \quad |\gamma| \ll 1. \label{eq:157} \\ \label{eq:139}
C_4[\beta]: x^{\mu} \to \hat{x}^{\mu} &=& [\,x^{\mu} + (x,x)\,\beta^{\mu}]/\sigma(x;\beta),\quad \mu = 0,1,2,3,
 \\  
&& \sigma(x;\beta) = 1 +2 (\beta,x) + (\beta,\beta)\,(x,x)\,; \nonumber \\ \label{eq:159}
 \delta x^{\mu} &=&  (x,x)\,\beta^{\mu}-2\,(\beta,x)\,x^{\mu},\,\, |\beta^{\mu}| \ll 1 .
\end{eqnarray}
From the infinitesimal scale transformation \er{eq:157} one obtains
\begin{equation}
  \label{eq:140}
  \partial_{\mu}s^{\mu}(x) = S^{\mu}_{\;\;\mu}, \quad s^{\mu}(x) = (\Theta^{\mu}_{\;\;\nu} + S^{\mu}_{\;\;\nu})x^{\nu}.
\end{equation}
And the four currents associated with the transformations \er{eq:159} obey the relations
\begin{equation}
  \label{eq:141}
  \partial_{\mu}k^{\mu}_{\,\,\rho}(x) = 2\,x_{\rho}\,S^{\mu}_{\;\;\mu}, \quad k^{\mu}_{\,\,\rho}(x) = 
(\Theta^{\mu}_{\;\;\nu} + S^{\mu}_{\;\;\nu})   [2\,x^{\nu}x_{\rho}
-(x,x)\delta^{\nu}_{\,\,\rho}],\,\, \rho = 0,1,2,3.
\end{equation}
Thus, for a vanishing electromagnetic current density $j^{\mu}=0\,(S_{\mu \nu} = 0)$ the five currents \er{eq:140} and 
\er{eq:141} are conserved, but
 for $j^{\mu} \neq 0$ this
is only the case if the trace $S^{\mu}_{\;\;\mu}$ vanishes. In general this does not happen! This was observed by 
Bessel-Hagen, too. Let me give two simple examples (with $c=1$):
For a charged relativistic point particle with (rest) mass $m$ one may take
\begin{equation}
  \label{eq:142}
  S^{\mu \nu}(x) = m\,\int_{-\infty}^{+\infty}d\tau\,\dot{z}^{\mu}\dot{z}^{\nu}\,\delta^4[x-z(\tau)],
\end{equation}
where $z(\tau)$ describes the orbit of the particle in Minkowski space.
 $S^{\mu \nu}(x)$ has the properties
\begin{equation}
  \label{eq:143}
  \partial_{\mu}S^{\mu \nu} = m \,\int_{-\infty}^{+\infty}d\tau\,\ddot{z}^{\nu}\,\delta^4[x-z(\tau)]= f^{\nu}(x),\quad
S^{\mu}_{\;\;\mu} = m \,\int_{-\infty}^{+\infty}d\tau\,\delta^4[x-z(\tau)].
\end{equation}
Thus, for a non-vanishing  mass the trace does not vanish and the five currents \er{eq:140} and \er{eq:141} are not
conserved. 

For a relativistic ideal fluid with invariant energy density $\epsilon (x)$ and invariant pressure $p(x)$ one has
\begin{equation}
  \label{eq:144}
  S^{\mu \nu}(x) = [\epsilon(x) + p(x)]u^{\mu}(x)u^{\nu}(x) - p(x) \eta^{\mu \nu},
\end{equation}
which has the trace
\begin{equation}
  \label{eq:145}
  S^{\mu}_{\;\;\mu} = \epsilon(x) - 3\,p(x), 
\end{equation}
which in general does not vanish either. It does so for a gas of massless particles and approximately so for
 massive particles
at extremely high energies. 

If the vector field $A_{\mu}(x)$ is coupled to a (conserved) current  $j^{\mu}(x)$ then the Lagrange density 
\begin{equation}
  \label{eq:163}
 L=  -\frac{1}{4}\,F_{\mu \nu}F^{\mu \nu}  - j^{\mu}A_{\mu}
\end{equation}
yields the field equations $\partial_{\mu}F^{\mu \nu} = j^{\nu}$ and the canonical energy-momentum tensor is now
\begin{equation}
  \label{eq:91}
   T^{\mu}_{\;\; \nu} = - F^{\mu \kappa}\partial_{\nu}A_{\kappa} + \delta^{\mu}_{\,\nu} (\frac{1}{4}\, F^{\kappa \lambda}
F_{\kappa \lambda} + j^{\mu}A_{\mu}). 
\end{equation} 
Its divergence is
\begin{equation}
  \label{eq:93}
  \partial_{\mu}T^{\mu}_{\;\;\nu}= (\partial_{\nu}j^{\lambda})A_{\lambda},
\end{equation}
i.e.\ an {\em external} current in general leads to ``violation'' of energy and momentum conservation  for the {\em
electromagnetic} subsystem. This is just a different version of Bessel-Hagen's analysis from above.

Nevertheless, the field equations $\partial_{\mu}F^{\mu \nu} = j^{\nu}$ and the expression
 $ j^{\mu}A_{\mu}\,dx^0dx^1dx^2dx^3$ are invariant
  under the transformations \er{eq:157} and \er{eq:159} if (see Eq.\ \er{eq:122})
\begin{eqnarray}
  \label{eq:162} \delta A_{\mu} &=& - \gamma\, A_{\mu}, \quad \delta j^{\mu} = -3\,\gamma \,j^{\mu}; \\
  \delta A_{\mu} &=& 2\,(\beta,x)\,A_{\mu} +2[\beta_{\mu}(x,A) - x_{\mu}(\beta,A)],  \label{eq:228} \\      
\delta j^{\mu} &=& 6\,(\beta,x)\,j^{\mu} +2[\beta^{\mu}(x,j) - x^{\mu}(\beta,j)], \nonumber
\end{eqnarray}
but this does not lead to new conservation laws if the current is an external one. Only if $j^{\mu}$ is composed of
dynamical fields, e.g.\ like the Dirac current $j^{\mu} = \bar{\psi}\gamma^{\mu}\psi$, conservation laws for the
total system may exist! (See also Subsect.\ 6.3).
\subsection{Invariances of an action integral versus  invariances of \\  associated differential equations
of motion}
A simple but illustrative example for the form invariance of an equation of motion without an additional conservation law
is the following:

A year before he presented E.\ Noether's paper to the G\"{o}ttingen Academy Felix Klein raised another in\-teresting
 question concerning ``dynamical'' differential equations, their symmetries and conservation laws: In 1916 Klein had
asked Friedrich Engel (1861-1941), a long-time collaborator of S.\ Lie, to derive the 10 known classical conservation laws
of the gravitational $n$-body problem by means of the 10-parameter Galilei Group, using group theoretical methods 
applied to  differential equations as developed by Lie. Engel did this by using Hamilton's equations and the invariance
properties of the canonical 1-form $ p_jdq^j - Hdt$ \cite{eng1}. Thus, he essentially already used the invariance of
 $Ldt\,$! \\
Then Klein noticed that the associated {\em equations of motion}, e.g.\
\begin{equation}
  \label{eq:146}
  m\,\frac{d^2\vec{x}}{d\,t^2} = -G\, \frac{\vec{x}}{r^3},\quad r= |\vec{x}|,
\end{equation}
 are also invariant under the transformation
 \begin{equation}
   \label{eq:147}
   \vec{x} \to \vec{x}^{\,\prime} = \lambda^2\,\vec{x}, \quad t \to t^{\prime} = \lambda^3\,t,\quad \lambda =
 \text{const.\ ,}
 \end{equation}
and he, therefore, asked Engel in 1917 whether this could yield a new conservation law. Engel's answer was negative 
\cite{eng2}. This had already been noticed in 1890 by Poincar\'{e} in his famous work on the 3-body problem \cite{poin}.

 Nevertheless, such joint scale transformations of space and time coordinates like \er{eq:147} may be quite useful, as
discussed by Landau and Lifshitz in their textbook on mechanics  \cite{lan}. In this context they also mention the 
virial theorem: If the potential $V(\vec{x})$ is homogeneous in $\vec{x}$ of degree $k$ and allows for bounded motions
such that $ |\vec{x}\cdot \vec{p}| < M < \infty\,$, then one gets for the {\em time averages} of the kinetic energy
$T= \vec{v}\cdot \vec{p}/2 = [d(\vec{x}\cdot \vec{p})/dt - \vec{x}\cdot\text{grad}V(\vec{x})]/2$ and the potential
 energy $V$:
\begin{equation}
  \label{eq:148}
  <T> = \frac{k\,E}{k+2}, \quad <V> = \frac{2\,E}{k+2}.
\end{equation}
These relations break down for $k= -2$, i.e.\ for potentials $V(\vec{x})$ homogeneous of degree $-2$. Here genuine scale
invariance comes in (not mentioned by Landau and Lifshitz!): For such a potential the expression 
$L\,dt = (T-V)\,dt$ is invariant with respect to the transformation
\begin{equation}
  \label{eq:149}
  \vec{x} \to \vec{x}^{\,\prime} = \lambda\,\vec{x}, \quad t\to t^{\prime} = \lambda^2\, t.
\end{equation}
This implies the conservation law
\begin{equation}
  \label{eq:150}
  S=   2\,E\,t - \vec{x}\cdot \vec{p}= \text{const.\ }.
\end{equation}
Thus, if $E \neq 0$, the term $\vec{x}\cdot \vec{p}= 2Et-S$ cannot be bounded in the course of time.

 For such a potential there is another conservation law, namely
\begin{equation}
  \label{eq:151}
  K = 2\, E\,t^2 - 2\,\vec{x}\cdot \vec{p}\,t + m\,\vec{x}^{\,2}= \text{const.\ }, 
\end{equation}
which, together with the constant of motion \er{eq:150}, determines $ r(t)$ without further integration.
The quantity $K$ can be derived, according to Noether's method, from the infinitesimal transformations
\begin{equation}
  \label{eq:152}
  \delta \vec{x} = 2\,\alpha\,\vec{x}\,t, \quad \delta t = 2\,\alpha\,t^2,\,\, |\alpha| \ll 1,
\end{equation}
which leave $L\,dt$ invariant up to a total derivative term $b\,dt,\,b= m\,d(\vec{x}^{\,2})/dt\,$.
Using Poisson brackets the three constants of motion $H=T+V,\,S$ and $K$ form the Lie algebra of the group
$SO(1,2)$ \cite{ka3a}.

That $n$-body potentials $V$ which are homogeneous of degree $-2$ with repect to their spatial coordinates have
 two additional conservation laws of the type \er{eq:150} and \er{eq:151} was already discussed by Aurel Wintner 
(1903-1958) in his impressive textbook \cite{win}, without, however, recognizing the group theoretical background.

Joint scale transformations of quantities with different physical dimensions and appropriate functions of them  was
in particular advocated by Lord Rayleigh (John William Strutt, 1842-1919) \cite{ray}.

\section{An arid period for conformal transformations \\ from about 1921 to about 1960}
From 1921 on conformal transformations and symmetries did not play a noticeable role in physics or mathematics. The
physics community was almost completely occupied with the new quantum mechanics and its consequences for atomic, molecular,
solid state, nuclear physics etc.. In mathematics there were several papers on the properties on conformal geometries as a 
consequence
of Weyl's work, mostly technically ones (see Ref.\ \cite{scho} for a long list of references), but also some as to 
classical field equations of physics. 

 With regard to the finite dimensional scale and special conformal transformations
there was Dirac's important paper from 1936 \cite{dir1} and -- independently -- about the same time the beginning of 
interpreting the special conformal ones as transformations from an inertial system to a system  which moves with a constant
acceleration with respect to the inertial one, an interpretation which ended in a dead end about 25 years later and brought
the group into discredit.
\subsection{Conformal invariance of classical field equations in physics}
In 1934 Schouten and Haantjes discussed conformal invariance of Maxwell's equations and of the associated 
 continuity equations for energy and
momentum  in the framework of Weyl's conformal geometry \cite{scho2}. 

 In 1935 Dirac proved invariance
 of Maxwell's equations with currents under the 15-parameter conformal group by rewriting them in terms of the
 hexaspherical coordinates
$y^0, \ldots,y^5,\,(y^0)^2 + (y^5)^2 - (y^1)^2 - \cdots - (y^4)^2=0,$ discussed in Subsect.\ 2.4 above. In addition
 he wrote down a spinor equation
\begin{equation}
  \label{eq:153}
  \beta^{\mu}\gamma^{\nu}\,M_{\mu \nu}\psi = \tilde{m}\, \psi, \quad M_{\mu \nu} = y_{\mu}\partial_{\nu} -
 y_{\nu}\partial_{\mu},
\,\, \mu,\nu =0,\ldots,5,
\end{equation}
where the $\beta^{\mu}$ and the $\gamma^{\nu}$ each are anticommuting $4 \times 4$ matrices, $\psi$ is a 4-component spinor
 and the 15
operators $M_{\mu \nu}$ correspond to the Lie algebra generators of the group $SO(2,4)$. The (dimensionless)
 number $\tilde{m}$ cannot be interpreted
as a mass (it is related to a Casimir invariant of $SO(2,4)$). Dirac tried to deduce from the spinor Eq.\ \er{eq:153}
  his original one, but did not succeed. This is no surprise: Eq.\ \er{eq:153} is invariant under the 2-fold covering
group $SU(2,2)$ of the identity component of $SO(2,4)$ the 15 generators of which can be expressed by the four Dirac
 matrices $\gamma^{\mu}, \mu = 0,1,2,3$ and their eleven independent products $\gamma^{\mu}\gamma^{ \nu}, \, \gamma^5 =
\gamma^0\gamma^1\gamma^2\gamma^3,\, \gamma^{\mu}\gamma^5.$ But in order to incorporate space reflections, one has to pass
to $8 \times 8$ matrices. The connection was later clarified by Hepner \cite{hep}.

 Also in  1935 Brauer and Weyl analysed spinor representations of pseudo-orthogonal groups in $n$ dimensions using
Clifford algebras and
clarified the topological structure of these groups for real indefinite quadratic forms \cite{brau}.

 Dirac's paper prompted only a few others at that time \cite{veb,bha,seg2}. 

 In 1936 Schouten and Haantjes showed
 \cite{scho3}, 
in the framework of Weyl's conformal geometry, that in addition to Maxwell's equations also the equations of motions
(geodetic equations in the presence of Lorentz forces) for a point particle are conformally invariant if one transforms
the  mass $m$ as an inverse length:
\begin{equation}
  \label{eq:154}
  m\,c/\hbar \to \omega^{-1/2}\,m\,c/\hbar,
\end{equation}
where $\omega$ is defined in Eq.\ \er{eq:104}. They also showed that the Dirac equation with non-vanishing  mass term
is invariant in the same framework, using  space-time dependent Dirac $\gamma$-matrices as introduced by Schr\"{o}dinger
and Valentin Bargmann (1908-1989)  in 1932 \cite{schr2} without mentioning the two.

 This was the first time that conformal invariance was enforced by transforming the mass parameter, too. It was used
and rediscovered frequently later on. Schouten and Haantjes did not discuss whether this {\em formal invariance} would
also imply a new conservation law as discussed by Bessel-Hagen!

In 1940 Haantjes discussed this mass transformation for the special conformal transformations \er{eq:139} applied to the
 relativistic
Lorentz force \cite{haa1}. He did the same about a year later for the usual Dirac equation with  mass term \cite{haa2}.

In 1940 Pauli argued that the Dirac equation as formulated for General Relativity by Schr\"{o}dinger and Bargmann could only
be conformally invariant in the sense of Weyl if the mass term vanishes and added in a footnote at the end that in any
conformally invariant theory the trace of the energy momentum tensor vanishes and that this could never happen for
systems with non-vanishing  masses \cite{pau}. 

 In 1956 Mclennan  discussed the conformal invariance and the associated conserved currents for free massless wave
equations with arbitrary spins \cite{mcl}.
\subsection{The acceleration ``aberration''}
Immediately after the papers by Bateman and Cunningham the conformal transformations were discussed
as a coordinate change between relatively accelerated systems by Hass\'{e}\cite{hass}. \\
Prompted by Arthur Milne's (1896-1950) controversial ``Kinematical Relativity'' \cite{miln}, Leigh Page (1884-1952) in 1935
proposed a ``New Relativity'' \cite{page} in which the registration of light signals should replace Einstein's rigid
measuring rods and periodical clocks. He came to the conclusion that in such a framework not only reference frames
which move with a constant relative velocity are equivalent, but also those which move with a relative constant
 acceleration! There was an immediate reaction to Page's papers by Howard Percy Robertson (1903-1961) \cite{rob1} who had
 already written critically before on Milne's work \cite{rob2}. Robertson argued that Page's framework should be 
accomodated within the kinematical one of General Relativity: He showed that the line element (Robertson-Walker)
\begin{equation}
  \label{eq:155}
  ds^2 = d\tau^2 - f^2(\tau)(dy_1^2 + dy_2^2 + dy_3^2),
\end{equation}
where $f(\tau)$ depends on a constant acceleration between observers Page had in mind, can be transformed into 
\begin{equation}
  \label{eq:156}
  ds^2 = [(t^2 -r^2/c^2)/t^2]^2\,[dt^2 - (dx_1^2 +dx_2^2+dx_3^2)/c^2].
\end{equation}
In addition he argued that the change of coordinate systems between observers with relative constant accelerations
 as described
 by Eq.\  \er{eq:155}
can be implemented by certain conformal transformations of the Minkowski-type line element \er{eq:156} which
Page had essentially used. 

Robertson further pointed out that the line element \er{eq:155} should better be related to a de Sitter universe.
He emphasized that the line element \er{eq:156}, too, should be interpreted in the kinematical framework of general
relativity and not in the one of special relativity, as Page had done. At the end he also warned that one should not
identify the conformal transformations he had used in connection with Eq.\ \er{eq:156} with the ones Bateman 
and Cunningham had discussed in the framework of special relativistic electrodynamics. 

Page hadn't mentioned conformal transformations at all and Robertson didn't mean to suggest that the transformations
his Ph.D. adviser Bateman had used in 1908/9 were to be interpreted as connecting systems of constant relative acceleration,
but the allusion got stuck and dominated the physical interpretation of the special conformal transformations for almost
thirty years.

 Page's attempt was also immediately interpreted in terms of the conformal group by Engstrom and Zorn, but without
 refering to accelerations \cite{eng}. 

 In 1940 Haantjes  also proposed \cite{haa1} to interpret the transformations \er{eq:139} as coordinate changes between
systems of constant relative acceleration, using hyperbolic motions \cite{paul2} without mentioning them explicitly:
Take $\beta = (0,b,0,0)$ and $x^0 = t,\,x^1 \equiv x, \,x^2 = x^3 = 0\,$. Then the spatial origin $(\hat{x} =0,\hat{x}^2 =0,
\hat{x}^3=0)$ of the new system moves in the original one according to
\begin{equation}
  \label{eq:158}
  0 = x(t) + b\,[t^2 -x^2(t)],\quad x^2 = 0,\quad x^3 =0,
\end{equation}
where
\begin{equation}
  \label{eq:160}
  \sigma(x;b) = 1-2b\,x(t) - b^2[t^2 - x^2(t)] = 1-b\,x(t)
\end{equation}
does not vanish  because $t^2 +1/(4b^2) > 0\,$. Haantjes does not mention Page nor Robertson.
 He summarizes his interpretation again in Ref.\ \cite{haa2}. \\
On the other hand, the time $\hat{t}$ of the moving system runs as
\begin{equation}
  \label{eq:199}
  \hat{t} = \frac{t}{1-b^2\,t^2}
\end{equation}
at the point $(x=0,\,x^2=0,\,x^3=0)\,$, i.e.\ it becomes singular for $(b\,t)^2 =1\,$. \\
The situation is similarly bewildering for the corresponding motion of a point $(\hat{x} =a \neq 0,\hat{x}^2 =0, 
\hat{x}^3=0)$ in the
original system: Instead of Eq.\ \er{eq:158}  we have now
\begin{equation}
  \label{eq:161}
  b\,(ab+1)[t^2-x^2(t)] +(2ab +1)\,x(t) - a = 0.
\end{equation}
As $b$ is arbitray, we may choose $ a\,b = -1$ and get from \er{eq:161} that $x= -a = 1/b$ and $\sigma(x;b) = -b^2\,t^2$.
If $a\,b = -1/2$, then \er{eq:161} takes the special form $t^2 - x^2(t) +1/b^2 =0$ and $\sigma(x;b) = 2(1-b\,x)$. \\
Many more such strange features may be added if one wants to keep the acceleration interpretation! The crucial point is
that the transformations \er{eq:139} are those of the Minkowski space and its associated inertial frames and
that one should find an interpretation within that framework. Accelerations are -- as Robertson asserted -- an element of
General Relativity! We come back to this important point below. 

 From 1945 till 1951 there were several papers by Hill
on the acceleration interpretation of the conformal group \cite{Hill} and about the same time work by Infeld and Schild
on kinematical cosmological models along the line of Robertson involving conformal transformations \cite{inf}.

 Then came a series of papers by Ingraham on conformal invariance of field equations, also adopting the acceleration
interpretation \cite{ing}. 

 Bludman, in the wake of the newly discovered parity violation, discussed conformal
 invariance of the 2-component neutrino equation and the associated $\gamma^5$-invariance of the massless Dirac equation
\cite{blud}, also mentioning the acceleration interpretation. 

 The elaborate final attempt to establish the conformal group as connecting reference systems with 
constant relative acceleration came from Rohrlich and collaborators \cite{rohr1}, till Rohrlich in 1963 conceded that the
 interpretation was untenable \cite{rohr2}!
\section{The advance of conformal symmetries \\ into  relativistic quantum field theories}
\subsection{Heisenberg's (unsuccesful) non-linear spinor theory \\ and  a few unexpected consequences}
Attempting to understand the mesonic air showers in cosmic rays, to find a  theoretical framework for the ongoing
 discoveries of new ``elementary'' particles, and to cure the infinities of relativistic non-linear quantum field
 theories, Werner 
Heisenberg (1901-1976) in the 1950s proposed a non-linear spinor theory as a possible ansatz \cite{heis1}. Spinors, because
one would like to generate half-integer and integer spin particle states, non-linear, because interactions among the 
basic dynamical quantum fields should be taken
into account on a more fundamental level, without starting from free particle field equations and inventing a
 quantum field theory for each newly discovered particle. The theory constituted a 4-fermion coupling on the
 Lagrangean level
which was not renormalizable according to the general wisdom. This should be taken care of by introducing a Hilbert space
with an indefinite metric (such introducing a plethora of new problems which were among the  reasons why the
theoretical physics community after a while rejected the theory). Heisenberg and collaborators associated the final
version
\begin{equation}
  \label{eq:164}
  \gamma^{\mu}\partial_{\mu}\psi \pm l^2\gamma^{\mu}\gamma^5\psi\,(\overline{\psi}\gamma_{\mu}\gamma^5\psi) = 0
\end{equation}
of their basic field equation with several symmetries \cite{heis2}, from which I mention  two which -- after detours -- had
 a lasting influence on future quantum field theories: 

In order to describe the isospin it was assumed that -- in analogy
to a ferromagnet --  the ground state carries an infinite isospin from which isospins of elementary particles emerge
 like spin waves.
This appears to be the first time that a degenerate ground state was introduced into a relativistic quantum field theory.
It was soon recognized by Nambu \cite{nam1} that the analogy to superconductivity was more fruitful for particle physics.

As spinors $\psi$ in the free Dirac equation have the dimension of length $[L^{-3/2}]$ ($\psi^+\psi$ is a spatial
 probability  density) the Eq.\ \er{eq:164} is invariant under the scale transformation \cite{heis2}
 \begin{equation}
   \label{eq:165}
   \psi(x,l) \to \psi^{\prime}(\rho\,x; \rho\,l) = \rho^{-3/2}\,\psi(x,l), \quad l \to l^{\prime}= \rho\,l, \quad \rho =
e^{\gamma}.
 \end{equation}
Here the length $l$ serves as a coupling constant which is not dimensionless. So Heisenberg et al.\ do the same what
Schouten and Haantjes had done previously \cite{scho3} with the mass parameter, namely to rescale it, too. As this does
not lead to a new conservation law the authors had to argue their way around that problem and they related possible 
associated discrete quantum numbers to the conservation of lepton numbers!

Though this attempt was not successful with respect to Eq.\ \er{eq:165}, it raised interest as to the possible role
of scale and conformal transformations in field theories and particle physics: Already early my later teacher Fritz
Bopp (1909-1987) had shown interest in them \cite{bopp}. Feza G\"{u}rsey (1921-1992) discussed the non-linear equation
\begin{equation}
  \label{eq:166}
  \gamma^{\mu}\partial_{\mu}\psi + \lambda \,(\overline{\psi}\,\psi)^{1/3}\,\psi = 0
\end{equation}
as an alternative which is genuine scale and conformal invariant \cite{guer}, though the cubic root is, of course, a
nuisance. McLennan added the more interesting example \cite{mcl2}
\begin{equation}
  \label{eq:167}
  \Box \vp + \lambda (\vp^*\vp)\,\vp = 0,
\end{equation}
where $\vp(x)$ is a complex scalar field in 4 dimensions with a dimension of lengh $[L^{-1}]\,$ and the coupling
constant $\lambda$ is dimensionless. 

Immediately after the paper by Heisenberg et al.\ with the scale transformation \er{eq:165} appeared, Julius Wess
(1934-2007) analysed
their interpretation of that transformation for the example of a free massive scalar quantized field, rescaling
 the mass parameter, too.
He found no conservation law in the massive case,
 but a time
 dependent generator for the scale transformation of the field \cite{wess1}.

 A year later  Wess published a paper \cite{wess2} in which he investigated the possible role of the conformal
 transformations
\er{eq:139} in quantum field theory, too, by discussing their role for free massless scalar, spin-one-half and
 electromagnetic
vector fields, including the associated conserved charges and the symmetrization of the canonical energy-momentum
tensor. He also analysed the conformal invariance of the 2-point functions. He further
  pointed out that the generator of scale transformations has a continuous spectrum and cannot provide discrete lepton
quantum numbers. He finally mentioned that in the case of a non-vanishing mass of the scalar field one can ensure 
invariance if
 one transforms the
mass accordingly. But no conservation law holds in that case. \\
Wess also observed that the transformations \er{eq:139} can map time-like Minkowski distances into space-like ones and
vice versa,  because
\begin{equation}
  \label{eq:168}
  (\hat{x}-\hat{y}, \hat{x}-\hat{y}) = \frac{1}{\sigma(x;\beta)\,\sigma(y; \beta)}\,(x-y,x-y),
\end{equation}
where 
\begin{equation}
  \label{eq:169}
  \sigma(x;\beta) = 1+ 2(\beta,x) + (\beta,\beta)\,(x,x) = (\beta,\beta)\,\left(x+\frac{\beta}{(\beta,\beta)}, x+
\frac{\beta}{(\beta,\beta)}\right).
\end{equation}
The sign of the product $\sigma(x;\beta)\,\sigma(y;\beta)$ may be negative! This mix-up of the causal structure for
the Minkowski space was, of course, a severe problem \cite{zee}. Similarly the possibility that $\sigma(x;\beta)$ from Eq.\ 
\er{eq:169} can vanish. At least locally causality is conserved because
\begin{equation}
  \label{eq:170}
  \eta_{\mu \nu}d\hat{x}^{\mu}\otimes d\hat{x}^{\nu} = \frac{1}{[\sigma(x;\beta)]^2}\,\eta_{\mu \nu}dx^{\mu}
 \otimes dx^{\nu}.
\end{equation}
Wess does not say anything about possible physical or geometrical interpretations of the conformal group!

\subsection{A personal interjection}
In 1959 I was a graduate student in theoretical physics and had to look for a topic of my diploma thesis. Being at the
University of Munich it was natural to go to Bopp, Sommerfeld's successor, who had been working on problems in quantum
mechanics and quantum field theory. He suggested a topic he was presently working on and which had to do with the fusion
of massless spin-one-half particles in an unconventional mathematical framework I did not like. Having learned from
talks in the nearby Max-Planck-Institute for Physics about scale transformations as treated in Ref.\ \cite{heis2}
 (Heisenberg and his Institute had moved from G\"{o}ttingen to Munich the year before), and knowing about Bopp's interest
in the conformal group, I asked him whether I could take that subject. Bopp was disappointed that I did not like his
original suggestion, but, being kind and conciliatory as usual, he agreed that I work on the conformal group!

When studying the associated literature, I got confused: whereas the interpretation of the scale transformations
 (dilatations) \er{eq:121} wasn't so controversial I couldn't make sense of the acceleration interpretation for the
conformal transformations \er{eq:139} (see Subsect.\ 5.2 above)! I knew I had to find a consistent interpretation in
order to think about possible physical applications. 

I was brought on the right track by the observation that there was a very close
 relationship between invariance or non-invariance of a system with respect to scale and conformal transformations:
If the dilatation current was conserved, so were the 4 conformal currents, if the dilatation current was not conserved,
then neither were the conformal ones, the divergences of the latter being proportional to the divergence of the former
one (see, e.g.\ Eqs.\ \er{eq:141}), at least in the cases I knew then. A simple example is given by a free relativistic
 particle \cite{ka4}: It follows from the
infinitesimal transformations \er{eq:157} and \er{eq:159} that the -- possibly -- conserved associated ``momenta'' are
given by ($c=1$)
\begin{eqnarray}
  \label{eq:171}
  s &=& (x,p) = E\,t -\vec{x}\cdot \vec{p},\quad E = (\vec{p}^{\,2} + m^2)^{1/2}. \\
 h^{\mu}&=& (x,x)\,p^{\mu}- 2\,(x,p)\,x^{\mu},\quad \mu = 0,1,2,3; \label{eq:172}  \\
h^0 &=& (t^2-r^2)\,E - 2\,s\,t, \;\; r=|\vec{x}|; \quad
\vec{h} = (t^2 -r^2)\,\vec{p} -2\,s\,\vec{x}. \label{eq:173}
\end{eqnarray}
Inserting
\begin{equation}
  \label{eq:174}
  \vec{x}(t) = (\vec{p}/E)\,t + \vec{a}
\end{equation}
into those momenta gives
\begin{eqnarray}
  \label{eq:175}
  s &=& -\vec{a}\cdot \vec{p} + (m^2/E)\,t, \\
h^0 &=& - \vec{a}^{\,2}\,E - (m^2/E)\,t^2, \quad 
\vec{h} = 2\,(\vec{a}\cdot\vec{p})\,\vec{a}- \vec{a}^{\,2}\,\vec{p} - (m^2/E)[(\vec{p}/E)\,t^2 + 2\,\vec{a}\,t],
 \label{eq:176} 
\end{eqnarray}
which shows that the quantities $s$ and $h^{\mu}$ are constants for a free relativistic particles only in the limits
$ m \to 0$ or $E \to \infty\,$! Both types are either conserved, or not conserved, simultaneously. 

More arguments for the conceptual affinities of scale and conformal transformations came from their group structures,
especially as subgroups of the 15-dimensional full conformal group. These features may be infered from the Lie algebra
 (now with hermitean generators, the Poincar\'{e} Lie algebra left out; compare also the algebra from Eqs.\ 
\er{eq:53} - \er{eq:59}, including the related group definitions of the operators):
\begin{eqnarray}
  \label{eq:178}
  [S,\,M_{\mu \nu}] &=& 0,\quad i\,[S,\,P_{\mu}] = -P_{\mu}, \quad i\,[S,\,K_{\mu}] = K_{\mu},\,\, \mu,
\nu = 0,1,2,3, \\
{[}K_{\mu},\,K_{\nu}{]} &=& 0, \label{eq:179} \\ i\,{[}K_{\mu},\,P_{\nu}{]} & =& 2\,(\eta_{\mu \nu}\,S - M_{\mu \nu}),
 \label{eq:229} \\  i\,
{[}M_{\lambda \mu}, \,
K_{\nu}{]} &=& (\eta_{\lambda \nu}\,K_{\mu} - \eta_{\mu \nu}\,K_{\lambda}). \label{eq:180}
\end{eqnarray}
These relations show that the transformations \er{eq:121} and \er{eq:139} combined form a subgroup (generated by $S$ and
$K_{\mu}$), that $K_{\mu}$ and $P_{\nu}$ combined do not form a subgroup, but generate scale transformations and homogeneous
Lorentz transformations. As
\begin{equation}
  \label{eq:225}
   K_{\mu} = R\cdot P_{\mu}\cdot R,
\end{equation}
 where $R$ is the inversion \er{eq:3},
one can generate the 15-dimensional conformal group by translations and the discrete operation $R\,$ alone \cite{ka1}!

Now all different pieces of the interpretation puzzle presented by the transformations \er{eq:139} fell into the right
places if one -- inspired by Weyl's conformal transformations \er{eq:104} and \er{eq:112} -- interpreted them as
 space-time dependent
scale transformations. However, whereas Weyl's factor $\omega(x)$ is arbitrary, the cor\-responding 
factor $1/\sigma^2(x; \beta)$
in Eq.\ \er{eq:170} has a special form induced by the coordinate transformations \er{eq:139}. For that reason I called
them ``special conformal transformations'' in Ref.\ \cite{ka1}, a name that has re\-mained.

 So the proposal was to interpret scale and special conformal transformations as (length) ``gauge transformations'' of the
Minkowski space \cite{ka2}, an interpretation which has been adopted generally by now.

Having a consistent interpretation did not immediately settle the question where those transformations could be physically
useful! A first indication came from the relations \er{eq:175} - \er{eq:176} which show that the very high energy limit
may be a possible realm for applications. It was helpful that -- at that time -- interesting interaction terms with
 dimensionless coupling constants like $\overline{\psi}\gamma^{\mu}\psi\,A_{\mu},\,\overline{\psi}\gamma^{5}\psi\,A,\,
\vp^4$ were also scale {\em and} conformal invariant \cite{ka4}. This led to Born approximations at very high energies and
very large momentum transfers which were compatible with scale invariance \cite{ka5}. However, the experimental hadronic
 elastic and other ``exclusive'' cross sections behaved quite differently. The way out was the deliberation that in these
reactions the scale invariant short distance properties were hidden behind the strong rearrangment effects of the long
range mesonic clouds which were accompanied by the emission of a large number of secondary particle into the final states,
like the emission oft ``soft'' photons in the scattering of charged particles.

A somewhat crude Bremsstrahlung model showed successfully how this mechanism could work and how to relate scale
 invariance to
the ``inclusive'' cross section (i.e.\ after summation over all final state channels) in inelastic electron-nucleon
scattering \cite{ka6}. A very similar result was obtained by Bjorken about the same time by impressively exploiting 
current algebra relations \cite{bjor1}. His scaling predictions for ``deep-inelastic'' electron-nucleon scattering
generated considerable general interest in the field \cite{bjor2}. Soon scale invariance at short distances found
 its proper place in
 applications of quantum field theories to high energy  problems  in elementary particle physics (see below).
 
The situation for special conformal transformations was more difficult at that time: First, there was their long bad
reputation of being related to a somewhat obscure coordinate change with respect to accelerated systems! I always felt the
associated resistance any time I gave a talk on my early work \cite{ka7}. Also, it appeared that scale invariance was
the dominating symmetry because special conformal invariance seemed to occur in the footsteps of scale invariance. This
changed drastically later, too. 
\subsection{Partially conserved dilatation and conformal currents, equal-time commutators and
 short-distance operator-expansions}
While I was in Princeton (University) in 1965/66, I was joined by the excellent student Gerhard Mack whom I had ``acquired''
in 1963 as my very first diploma student in Munich. In Princeton it was not easy to persuade Robert Dicke (1916-1997)
 who was in 
charge of admissions that Mack would be an adequate graduate student of the physics department, but I succeeded! Around that
time the work on physical consequences from Murray Gell-Mann's algebra of currents was at the forefront of activities
in theoretical particle physics \cite{curr1,curr2}. In discussions with John Cornwall who had invited me for a fortnight to
UC Los Angeles to talk about my work, the idea came up to incorporate broken scale and conformal invariance
into the current algebra framework. Back in Princeton I suggested to Gerhard Mack to look into the problem. This he did
with highly impressive success: When we both were in Bern in 1966/67 as guests of the Institute for Theoretical Physics 
(the invitation arranged by Heinrich Leutwyler), he completed his Ph.D.\ thesis on the subject \cite{mac2}
and got the degree in Februar 1967 from the University of Bern. An extract of the thesis was published in 1968 \cite{mac3}.

In the next few years the subject almost ``exploded''. It is impossible to cover the different lines of development in these
brief notes
and I refer to several of the numerous reviews
 \cite{mac1,gell1,gat,baru1,zu,carr,col,jac,gat1,rue1,tod3,mac8,gat2,mac4,tod1,mac5,pol,ket,fra,nahm1} on
the field. I here shall briefly indicate only the most salient steps till today:

The crucial new parameter which is associated with scale (and special conformal) transformations is the length dimension
$l$ of a physical quantity $A$: It is said to have the length dimension $l_A$ if it transforms under the group \er{eq:121}
as \cite{ka1}
\begin{equation}
  \label{eq:182}
  S_1[\gamma]\,: \quad A \to \hat{A} = \rho^{l_A}\,A, \quad \rho = e^{\gamma};
\end{equation}
or, if $F(x)$ is a field variable,
\begin{equation}
  \label{eq:183}
  F(x) \to \hat{F}(\hat{x}) = \rho^{l_F}\,F(x), \quad \hat{x} = \rho\,x.
\end{equation}
If $A$ or $F(x)$ are corresponding operators and scale invariance holds, then 
\begin{equation}
  \label{eq:184}
  e^{i\,\gamma\,S}\,A\,e^{-i\,\gamma\,S} = \rho^{l_A}\,A, \quad e^{i\,\gamma\,S}\,F(x)\,e^{-i\,\gamma\,S}= 
\rho^{-l_F}\,F(\rho\,x),
\end{equation}
where $S$ is the hermitean  generator of the scale transformation (= dilatation). 
The ``infinitesimal'' versions of these relations are
\begin{equation}
  \label{eq:185}
  i\,[S,\,A] = l_A\,A, \quad i\,[S,\,F(x)] = (-l_F + x^{\mu}\partial_{\mu})\,F(x).
\end{equation}
The corresponding relations for the generators $K_{\mu}$ of the special conformal transformations are
\begin{equation}
  \label{eq:186}
  i\,[K_{\mu},\,F(x)] = [-2\,x_{\mu}(-l_F +x^{\nu}\partial_{\nu}) + (x,x)\partial_{\mu} +2\,x^{\nu}\Sigma_{\mu \nu}]\,F(x),
\end{equation}
where the $\Sigma_{\mu \nu}$ are the spin representation matrices of $F$ with respect to the Lorentz group.
The ``classical'' or ``canonical'' dimension of scalar and vector fields $\vp(x)$ and $A_{\mu}(x)$ in 4 space-time
 dimensions
are $l_{\vp} = l_A = -1$ (the classical action integral $\int d^4x\, L$ has vanishing length dimension), a Dirac spinor
$\psi(x)$ has $ l_{\psi} =-3/2$. \\
It follows from the commutation relations \er{eq:178} that $M_{\mu \nu},\,P_{\mu}$ and $K_{\mu}$ have the dimensions
$0,\,-1$ and $+1$, respectively. As a mass parameter $m$ has -- in natural units (see Eq.\ \er{eq:154}) -- length
dimension $-1$, one defines the ``mass dimension'' $d_F = -l_F$, in order to avoid the minus signs in case of the usual
 fields.

For a given scale invariant system one expects the generator $S$ to be the space integral
\begin{equation}
  \label{eq:187}
  S = \int dx^1dx^2dx^3\,s^0(x)
\end{equation}
 of the component $s^0$ of
the dilatation current $s^{\mu}(x)$, where -- according to Eq.\ \er{eq:118} --
\begin{equation}
  \label{eq:188}
  s^{\mu}(x) = T^{\mu}_{\;\;\nu}\,x^{\nu}+ \sum_i d_i\frac{\partial L}{\partial(\partial_{\mu}\vp^i)}\vp^i.
\end{equation}
Using the equations of motions $E(\vp^i) = 0$ (Eq.\ \er{eq:127}) and the expression \er{eq:128} for the canonical
energy-momentum tensor, we get for the divergence
\begin{eqnarray}
  \label{eq:189}
  \partial_{\mu}s^{\mu}& =& T^{\mu}_{\;\; \mu} + \sum_id_i\,\frac{\partial L}{\partial \vp^i}\,\vp^i + d_i\,\frac{\partial L}{
\partial(\partial_{\mu} \vp^i)}\,\partial_{\mu}\vp^i \\ & =& - 4\,L + 
\sum_id_i\,\vp^i\,\frac{\partial L}{\partial \vp^i} + (d_i+1)\, \partial_{\mu}\vp^i\, \frac{\partial L}{
\partial(\partial_{\mu} \vp^i)}. \nonumber
\end{eqnarray}
For a large class of models the divergence of the special conformal currents is proportional to the divergence \er{eq:189},
namely
\begin{equation}
  \label{eq:197}
   \partial_{\mu}k^{\mu}_{\;\; \lambda}(x) = 2\,x_{\lambda}\,\partial_{\mu}s^{\mu}(x),\quad \lambda = 0,1,2,3.
\end{equation}
For this and possible exceptions see Refs.\ \cite{mac1,wess3,pol}. Compare also Eq.\ \er{eq:141} above.
 \\
If the divergence $\partial_{\mu}s^{\mu}$ vanishes then the generator $S$ from Eq.\ \er{eq:187} is -- formally at
 least -- independent of time and
it follows from the commutation relations \er{eq:178} that the mass squared operator obeys
\begin{equation}
  \label{eq:190}
  M^2 = P_{\mu}P^{\mu} \to \hat{M}^2 = e^{i\,\gamma\,S}\,M^2\,e^{-i\gamma\,S} = \rho^{-2}\,M^2,
\end{equation}
i.e.\ either $M^2 =0$ or $M^2$ has a continuous spectrum. In general, however, the physical spectrum of $M^2$ is more
 complicated
and the divergence \er{eq:189} will not vanish. A very simple example is provided by a free massive scalar field which
has (ignoring problems as to the product of field operators at the same point in the quantum version!)
\begin{equation}
  \label{eq:191}
  L = \frac{1}{2}(\partial_{\mu}\vp\,\partial^{\mu}\vp -m^2\,\vp^2), \quad \partial_{\mu}s^{\mu} = m^2\,\vp^2.
\end{equation}
(As to problems with the energy-momentum tensor for quantized scalar fields see Refs.\ \cite{curr2,col}, the references
 given
there and the literature quoted in Subsect.\ 6.4 below \cite{anom}.) 

In such a case  as \er{eq:191} the generator \er{eq:187} becomes time dependent: $S=S(x^0)$. Nevertheless, using
 canonical equal-time
commutation relations for the basic field variable and their conjugate momenta the second of the relations \er{eq:185}
may still hold (formally) and can be exploited, mostly combined with the fact that the divergence is a local field 
with spin $0$, carrying certain internal quantum numbers \cite{mac3,gell1,gat,ell}. In this approach the dimensions
 $d_i$ of the
fields $\vp^i$ are considered to be the prescribed canonical ones. 

Postulating the existence of equal-time commutators is seriously not tenable in the case of interacting fields which
may also acquire non-canonical dimensions. This was first clearly analyzed by Kenneth Wilson in the framework of his
operator product expansion \cite{wil1}: In quantum field theories the product of two ``operators'' $A(x)$ and $B(y)$
is either badly defined or highly singular in the limit $y \to x$. Refering to experience with free fields and perturbation
theory Wilson first postulated that
\begin{equation}
  \label{eq:192}
  \text{for }\,\,\, y \to x: \quad A(x)\,B(y) \simeq \sum_n C_n(x-y)\,O_n(x),
\end{equation}
where the generalized functions $C_n(x-y)$ beome singular on the light cone $(x-y)^2 = 0$ and the operators $O_n(x)$
 are local fields.

 Second Wilson implemented asymptotic scale invariance by attributing (mass) dimensions $d_A,\, d_B$ and $d_n$ to 
the fields in Eq.\ \er{eq:192}. Postulating the property \er{eq:184} for both sides of that (asymptotic) relation
implies
\begin{equation}
  \label{eq:193}
  C_n[\rho\,(x-y)] = \rho^{d_n -d_A -d_B}\,C_n(x-y),
\end{equation}
i.e.\ the $C_n(x)$ are homogeneous ``functions'' of degree $d_n -d_A -d_B$. \\ This approach provided an important
operational framework for the concept ``asymptotic scale invariance''. It was soon applied successfully to deep inelastic
lepton-hadron scattering processes, with $A$ and $B$ electromagnetic current operators $j_{\mu}(x)$ of hadrons
 \cite{cicc,gat,mac6,carr,deep}.
\subsection{Anomalous dimensions,  Callan-Symanzik equations, conformal anomalies  and conformally
 invariant $n$-point functions}

Wilson had already stated that for interacting fields the dimensions $d_A,\,d_B$ and $d_n$ need not be canonical. The
intuitive reason for this is the following: even if one starts with a (classical) massless scale invariant theory in
lowest order, one (always) has to break the symmetry in higher order perturbation theory for renormalizable field theories
 through the introduction of regularization schemes which involve length (mass) parameters like a cutoff, e.g.\ related
 to a mass (re)normalization point $\mu$. 
This leads to anomalous positive corrections to the canonical dimensions 
\begin{equation}
  \label{eq:195}
  d \to d + \gamma(g),\,\,\gamma(g) \geq 0,
\end{equation}
 where $g$ is the physical coupling constant. 

 Curtis Callan and Kurt Symanzik (1923-1983) in 1970 showed
 (independently) how a change of the
scale parameter $\mu$ and an associated one in the coupling $g$ affects the asymptotic behaviour of the $n$th proper
 (1-particle irreducible) renormalized Green function in momentum space \cite{sym}. For a single scalar field its 
asymptotic behaviour is governed by
the partial differential (Callan-Symanzik) equation
 \begin{equation}
   \label{eq:194}
  \left(\mu\, \frac{\partial}{\partial \mu} + \beta(g)\,\frac{\partial}{\partial g} - n\,\gamma(g)\right)\Gamma^{(n)}_{as}
(p; g, \mu) = 0.
 \end{equation}
The function $\beta(g)$ and the anomalous part $\gamma(g)$ of the dimension may be calculated in perturbation theory.
 This material is discussed in standard textbooks \cite{itz,zin,wei1} and therefore I shall not discuss it further here.

As there is no regularization scheme which preserves scale and special conformal invariance, the quantum version of
the trace of the energy momentum tensors and the divergences of the dilatation and special conformal currents in general
contain an anomaly which for a number of important models is proportional to $\beta(g)\,$, where $\beta(g)$ is the same
 function
as in Eq.\ \er{eq:194} \cite{anom}. Thus, only if $\beta(g)$ vanishes identically or has a zero (fix point) at some
 $g=g_1 \neq 0$ can
the quantum version of that field theory be dilatation and conformally invariant. The vanishing of $\beta(g)$ in all orders
of perturbation theory, and perhaps beyond, appears to happen for the $\mathcal{N} = 4$ superconformal pure
Yang-Mills theory (see Subsect.\ 7.3 below).

Early it seemed, at least in Lagrangean quantum field theory \cite{mac1,wess3}, that in most cases scale invariance
entails special conformal invariance, e.g.\ the 2-point function 
\begin{equation}
  \label{eq:196}
  \langle 0|\vp(x)\,\vp(0)|0 \rangle = \text{const. } (x^2-i0)^{-d}
\end{equation}
of a scale invariant system with scalar field $\vp(x)$  is also conformally invariant. Here the dimension $d$ 
of the scalar field in general remains to be determined by  the interactions.

 But then in 1970 Schreier \cite{schrei}, Polyakov \cite{poly} and Migdal \cite{mig1} realized that
conformal invariance imposes {\em  additional} restrictions on the 3-point functions, leaving only a few parameters
to be determined by the dynamics.

Migdal also proposed a self-consistent scheme (``bootstrap'') in terms of Dyson-Schwinger equations with
 Bethe-Salpeter-like kernels which, in principle,
would allow to find solutions for the theory. This soon led to  an impressive development in the analysis of
 conformally invariant 3- and 4-dimensional quantum field theories in terms of (euclidean) $n$-point functions,
 reviews of which can be found  in Refs.\ \cite{rue1,tod3,mac8,mac4,tod1}.
\section{Conformal quantum field theories in 2 dimensions, global properties of conformal
 transformations, \\ supersymmetric conformally invariant systems, \\ and ``postmodern'' developments}
\subsection{2-dimensional conformal field theories}
Polyakov's paper \cite{poly} was a first important step for conformal invariance into the realm of statistical physics,
but progress was somewhat slow due to the fact that the additional symmetry group was just 3-dimensional (scale
 transformations
were already established within the theory of critical points in 2nd order phase transitions). However applications of
conformal invariance again ``exploded'' after the seminal paper \cite{poly2}  by Belavin, Polyakov and Zamolodchikov on
 {\em 2-dimensional} conformal invariant quantum field theories with  now ``infinite-dimensional'' conformal Lie
algebras generated by the  (Witt) operators
\begin{equation}
  \label{eq:198}
  l_n = z^{n+1}\frac{d}{dz}, \quad n= 0,\,\pm 1,\, \pm 2, \, \ldots, \quad [l_m,\, l_n] = (m-n)\,l_{m+n},
\end{equation}
if one uses complex variables, and the corresponding  ones with  complex conjugate variables $\bar{z}$ where 
 $\bar{l}_n = \bar{z}^{n+1}d/d\bar{z}$.
 The generators \er{eq:49} - \er{eq:52} form a (real) 6-dimensionsional subalgebra which generates the group \er{eq:38}. 
Its complex basis here is $l_{-1},\,l_0,\,l_1$ and $\bar{l}_{-1},\,\bar{l}_0,\,\bar{l}_1$. 

The ``quantized'' version of the algebra \er{eq:198} generated by corresponding operators $L_n$,
 the Virasoro algebra, contains anomalies which can be interpreted as those of the 2-dimensional energy-momentum tensor
 which, being conserved and having a vanishing trace, has only 2 independent components one of which can be (Laurent)
 expanded in terms of the
 $L_n$, the other
one in terms of the $\bar{L}_n$ \cite{vir}. \\
The Virasoro algebra first played an important role for the euclidean version of the 2-dimensional bosonic string world
 sheet \cite{gre,pol2,nahm1}. However, with the work of Belavin et al.\ \cite{poly2} 2-dimensional conformal quantum field
theory became a subject by its own, first, because it allowed rich new insights into the intricate structures of such
 theories
\cite{mac5,ket,nahm1,gaw}, and second, because it allowed for applications in statistical physics for, e.g.\ phase
 transitions in
  2-dimensional surfaces \cite{itzy,gin,chr,ket,card}.
\subsection{Global properties of conformal transformations}
The transformations \er{eq:3} and \er{eq:139} become singular on light cones. As was already known in the 19th century,
this can be taken care of geometrically in terms of the polyspherical coordinates discussed in Subsect.\ 2.4 above which
allow to extend the usual Minkowski space and have all conformal transformations act linearly on the extension. 
One  can even give  a physical interpretation of the procedure \cite{ka1,ka2}: Introducing homogeneous coordinates
\begin{equation}
  \label{eq:200}
  x^{\mu} = y^{\mu}/k, \quad \mu = 0,\,1,\,2,\,3,
\end{equation}
one can interpret $k$ as an initially Poincar\'{e} invariant length scale which transforms as
\begin{eqnarray}
  \label{eq:201}
  k \to \hat{k} &=& e^{-\gamma}\,k, \\
 k \to \hat{k} &=& \sigma(x;\beta)\,k, \label{eq:202}
\end{eqnarray}
with respect to the groups \er{eq:121} and \er{eq:139}. The limit $\sigma(x;\beta) \to 0$ then means that at the points
with the associated coordinates $x$ the new scale $\hat{k}$ becomes infinitesimally small so that the new coordinates
 $\hat{x}^{\mu}$
become arbitrarily large whereas the dimensionless coordinates $y^{\mu}$ stay the same \cite{ka1,ka2}. The scale
coordinate $\hat{k}$ can
 also become negative now. 

In order to complete the picture, one introduces the dependent coordinate
\begin{equation}
  \label{eq:203}
  q = (x,x)\,k, \quad \text{ or } \quad Q(y;q,k) \equiv (y,y) -q\,k = 0,
\end{equation}
where q has the dimension of length, transforms as
\begin{equation}
  \label{eq:204}
  q \to \hat{q} = e^{\gamma}\,q
\end{equation}
under dilatations and remains invariant under special conformal transformations. So we have now
\begin{eqnarray}
  \label{eq:205}
  C_4[\beta]\,:\quad \hat{y}^{\mu} & =& y^{\mu} +\beta^{\mu}\,q, \\
\hat{k} &=& 2\,(\beta,y) + k + (\beta,\beta)\,q, \nonumber \\
\hat{q} &=& q. \nonumber
\end{eqnarray}
These transformations leave the quadratic form $Q(y;q,k)$ itself invariant, not only the equation $Q=0$.
The same holds for the translations $T_4[a]: x^{\mu} \to x^{\mu} + a^{\mu}$ which act on the new coordinates as
\begin{eqnarray}
  \label{eq:206}
  T_4[a]\,: \quad \hat{y}^{\mu} &=& y^{\mu} + a^{\mu}\,k, \\
\hat{k} &=& k, \nonumber \\
\hat{q} &=& 2\,(\beta,y) + (\beta,\beta)\,k +q. \nonumber
\end{eqnarray}
It is important to keep in mind that a given 6-tuple $(y^0,\,y^1,\,y^2,y^3,\,k,\,q)$, with $Q(y;k,q) =0$, is only one
representative of an equivalence class of such 6-tuples which can differ by an arbitrary real multiplicative number
 $\tau \neq 0\,$
and still describe the same point in the 4-dimensional physical space:
\begin{equation}
  \label{eq:207}
  (y^0,\,y^1,\,y^2,\,y^3,\,k,\,q) \cong \tau \, (y^0,\,y^1,\,y^2,\,y^3,\,k,\,q), \quad \tau \neq 0.
\end{equation}
As $k$ is a Lorentz scalar, the quadratic form $(y,y)$ is -- like $(x,x)$ -- invariant under the  homogenous Lorentz
 group $O(1,3)$. So 
the 15-parameter
 conformal group $C_{15}$ of the Minkowski space $M^4$ leaves the quadratic form $Q(y;k,q)$ invariant. Writing
\begin{eqnarray}
  \label{eq:208}
  k&=& y^4 + y^5,\quad q = y^4 - y^5, \\ \quad Q(y,y) &\equiv& (y^0)^2 + (y^5)^2 - (y^1)^2 - (y^2)^2 - (y^3)^2
 - (y^4)^2 \label{eq:213} \\ &=&  y^T\cdot\eta\cdot y =
\eta_{i j}y^i\,y^j, \nonumber
\end{eqnarray}
one sees how the group $C_{15}$ has to be related to the group $O(2,4)\,$. Because of the equivalence relation \er{eq:207}
one has
\begin{equation}
  \label{eq:209}
  C_{15} \cong O(2,4)/Z_2(6), \quad Z_2(6): y \to y, \text{ and } y \to -y.
\end{equation}
$Z_2(6)$ is the center of the identity component $SO^{\uparrow}(2,4)$ of $O(2,4)$ to which $C_4[\beta]$ and
$S_1[\gamma]$ belong, too. 

Like $O(1,3)$ the group $O(2,4)$ -- and therefore $C_{15}$, too -- consists of 4 disjoint pieces: If $w \in O(2,4)$ is
 the transformation matrix defined by
\begin{equation}
  \label{eq:214}
  y^i \to \hat{y}^i = w^i_{\,\,j}\,y^j,\quad i=0,\,\ldots,\,5,\quad w^T\cdot\eta\cdot w = \eta,
\end{equation}
then the pieces are characterized by \cite{brau}
\begin{equation}
  \label{eq:215}
  \det w = \pm 1, \quad \epsilon_{0 5}(w) \equiv \text{sign}(w^0_{\,\,0}w^5_{\,\,5}-w^0_{\,\,5}w^5_{\,\,0}) = \pm 1,
\end{equation}
the identity component $SO^{\uparrow}(2,4)$ being given by $\det w =1,\,\epsilon_{05}(w) =1\,$.
For, e.g.\ the inversion \er{eq:3} we have 
\begin{equation}
  \label{eq:216}
  R\,: \quad k \to q,\,q \to k,\,\, \text{ or }\,\, y^0 \to y^0,\cdots, \,y^4 \to y^4,\,y^5 \to -y^5;\,\,
\det w = -1,\,\,
\epsilon_{0 5}(w) = -1.
\end{equation}
 For these and other details I refer to the literature
 \cite{go,go1,mac7,tod2,mac5}. 

 Taking the equivalence relation \er{eq:207} into account, one can express $Q(y,y) = 0$ as
\begin{equation}
  \label{eq:210}
  (y^0)^2 + (y^5)^2 = (y^1)^2 +(y^2)^2 + (y^3)^2 + (y^4)^2 = 1,
\end{equation}
and one sees that the extended Minkowski space on which the group $C_{15}$ acts continuously is compact and
 topologically given by
\begin{equation}
  \label{eq:211}
  M^4_c \simeq (S^1 \times S^3)/Z_2(6),
\end{equation}
which  essentially says that time is compactified to $S^1$ and space to $S^3\,$. So time becomes periodic!
 Now one
needs at least four coordinate neighbourhoods (``charts'') in order to cover the manifold \er{eq:211} \cite{go,go1}.
 This  can be related
to the fact that the maximally compact subgroup of the identity component of $O(2,4)$ is $SO(2) \times SO(4)\,$.

 In order to get rid of the periodical time one has to pass to the universal covering space
\begin{equation}
  \label{eq:212}
  M^4_c \to \tilde{M}^4 \simeq \mathbb{R} \times S^3.
\end{equation}
If $\tl{M}^4$ is parametrized as
\begin{equation}
  \label{eq:218}
  \tl{M}^4 = \{ (\tau, \vec{e});\, \tau \in \mathbb{R},\, \vec{e}= (e^1,\,e^2,\,e^3,e^4) \in S^3,\,\vec{e}^{\,2}=1 \},
\end{equation}
then that coordinate chart for $M^4_c$ which describes the relations \er{eq:210} for $k=y^0+y^5 > 0\,$, namely $M^4$,
 may be given by
\begin{equation}
  \label{eq:219}
  y^0 = \sin \tau,\quad y^5 = \cos \tau,\quad \tau \in (-\pi,\,+\pi),\quad \vec{y} = \vec{e},\quad e^4 + \cos \tau > 0.
\end{equation} leading to the Minkowski space coordinates 
\begin{equation}
  \label{eq:220}
  x^0 = \frac{\sin \tau}{e^4 + \cos \tau}, \quad x^j = \frac{e^j}{e^4 + \cos \tau},\quad j=1,\,2,\,3,
\end{equation}
from which it follows that
\begin{equation}
  \label{eq:226}
  \eta_{\mu \nu}\,dx^{\mu}\otimes dx^{\nu} = \frac{1}{(e^4 + \cos \tau)^2}\,(d\tau \otimes d\tau - \sum_{i=1}^4
de^i \otimes de^i), \quad \vec{e}\cdot d\vec{e} = 0,
\end{equation}
or more generally -- according to Eqs.\ \er{eq:200}, \er{eq:208} and \er{eq:213} --
\begin{equation}
  \label{eq:227}
  \eta_{\mu \nu}\,dx^{\mu}\otimes dx^{\nu} = \frac{1}{k^2}\,(\eta_{ij}\,dy^i \otimes dy^j), \quad y_jdy^j = 0.
\end{equation}
 Global structures associated with conformal point transformations were first analyzed by Kuiper \cite{kui}.

Very important in this context is the question of possible causal structures on those manifolds, i.e.\ whether one can give
a conformally invariant meaning to time-like, space-like and light-like. We know already that such a global causal structure
 cannot exist
 on $M^4\,$ (Eq.\ \er{eq:168}),
but that a local one is possible (Eqs.\ \er{eq:170} and \er{eq:227}). This  generalizes to the compact
 manifold $M^4_c$ \cite{go}.

 However, the universal covering $\tilde{M}^4$ does allow for a {\em global conformally invariant
 causal
 structure}  with respect to the universal covering group $\widetilde{SO^{\uparrow}(2,4)}\,$,
namely a point $(\tau_2,\vec{e}_2) \in \tl{M}^4$ is {\em time-like later} than $(\tau_1, \vec{e}_1)$ if
\begin{equation}
  \label{eq:221}
  \tau_2-\tau_1 > \text{Arccos}(\vec{e}_1\cdot \vec{e}_2),
\end{equation}
where $y= \text{Arccos}(x)$ means the principal value $y \in [0,\,\pi]$.
More in the Refs.\  \cite{seg,go,go1,mac7,tod2,mac5}. That the universal covering space \er{eq:218} allows for a
conformally invariant structure was first observed by Segal \cite{seg}.

Now the group $O(2,4)$ is also the invariance group (``group of motions'') of the anti-de Sitter space
\begin{equation}
  \label{eq:217}
  AdS_5\,: \quad Q(u, u) = (u^0)^2 + (u^5)^2 - (u^1)^2 - (u^2)^2 - (u^3)^2 - (u^4)^2 = a^2,
\end{equation}
which has the topological structure $S^1 \times \mathbb{R}^4\,$ and is, therefore, multiply connected, too. It has the
 universal
covering $\mathbb{R} \times \mathbb{R}^4\,$ or $\mathbb{R} \times S^4\,$ if one compactifies the ``spatial'' part
 \cite{haw}. 

In the following sense the  Minkowski space \er{eq:220} may be interpreted as part of the boundary of the space \er{eq:217}
 \cite{witt,rehrev}: Introducing the coordinates
\begin{equation}
  \label{eq:222}
  u^0 = (a^2 + \lambda^2)^{1/2}\,\sin \tau,\quad u^5 = (a^2 + \lambda^2)^{1/2}\,\cos \tau,\quad \vec{u}=
\lambda\,\vec{e},\quad \lambda >0,\quad \vec{e} \in S^3,
\end{equation}
the ratios
\begin{equation}
  \label{eq:223}
  \xi^0=\frac{u^0}{u^4+u^5},\quad \xi^j = \frac{u^j}{u^4+u^5}, \quad j=1,\,2,\,3,
\end{equation}
approach the limits \er{eq:220} if $\lambda \to \infty\,$. Such limits may be taken for other charts of the compact space
\er{eq:211}, too. 

This property is at the heart of tremendous activities in
a part of the mathematical physics community during the last decade (see the last Subsect.\ below).

As the group $O(2,4)$ is infinitely connected (because it contains the compact subgroup $SO(2) \cong S^1$) it has also
infinitely many covering groups, the double covering $SU(2,2)$ being one of the more important ones. The theory of
irreducible representations of $O(2,4)$ and its covering groups also belongs to its global aspects. I here mention
 only a few of the relevant references on the discussions of those irreducible representations \cite{irr}.
\subsection{Supersymmetry and conformal invariance}
An essential key to the compatibility of conformal invariance and supersymmetries is the generalization of the
 Coleman-Mandula theorem \cite{col3} by Haag, {\L}opusz\'{a}nski and Sohnius \cite{haag}: Coleman and Mandula had
 settled a long
dispute about the possible non-trivial ``fusion'' of space-time (Poincar\'{e}) and internal symmetries. From reasonable
assumptions they deduced that one can only have a non-trivial $S$-matrix, if Poincar\'{e} and internal symmetry group
``decouple'', i.e.\ form merely a direct product. One of their postulates was the existence of a finite mass gap, thus
excluding conformal invariance. Soon afterward supersymmetries were discovered \cite{soh,west3,wess4,west1,wein2,witt} the
 fermionic
generators of which had non-trivial commutators with those of the Poincar\'{e} group. Haag et al.\ not only generalized
Coleman's and Mandula's results in the massive case by including supersymmetric charges, but they also discussed the
case of massless particles and found a unique structure: now the fermionic supercharges can generate the 15-dimensional 
Lie algebra of the conformal group {\em and}  internal unitary symmetries $U(\mathcal{N}),\,\mathcal{N} = 1,\,2, \ldots,
\,8\,$. For their result the inclusion of the generators $K_{\mu}$ of the special conformal transformations was essential. 

One very important feature of supersymmetries is that they reduce the number of divergences for the conventional
 quantum field
theories, making the usual associated renormalization procedures at least partially superfluous (so-called
 ``non-renormalization theorems'')
\cite{soh,west3,wess4,west1,wein2}. Of special interest here are the $\mathcal{N}= 4$ superconformal quantum Yang-Mills
 theories in 4 space-time dimensions : they are finite, their $\beta$-function (see Subsect.\ 6.4 above) vanishes 
\cite{soh1,wein3,west2,witt} and, therefore, the trace
of their energy-momentum tensor, too, thus implying scale and conformal invariance on the quantum level! The Lorentz spin
 content of this system of massless fields is:
1 (Yang-Mills) vector field with two helicities and an $SU(N)$ gauge group, $4$ spin $1/2$ fields with two helicities each
and 6 scalar fields. 

  Though this model is as similar to
physical realities as an ideal mathematical sphere is similar to the real earth with its mountains, oceans, forests,
towns etc., it is nevertheless a striking and interesting idealization! 
\subsection{``Postmodern'' developments}
\subsubsection{AdS/CFT correspondence}
The last 10 years have seen thousands (!) of papers which are centered around a conjecture by Maldacena \cite{mal} related
 to the limit $\lambda
\to \infty$ of Eq.\ \er{eq:223} above, namely that the 4-dimensional conformally compactified Minkowski space (or its
universal covering) is the
 boundary of  the 5-dimensional anti-de Sitter space $AdS_5$ (or its universal covering correspondingly). 
The conjecture is that the superconformal $\mathcal{N}
=4,\, SU(N)$ Yang-Mills theory on Minkowski space  ``corresponds'' (at least in the limit
 $N \to \infty$) to a (weakly coupled) supergravity
 theory on
$AdS_5$ accompanied by a Kaluza-Klein factor $S^5\,$,  both factors being related to a superstring in 10 dimensions of type
 $II\,B$
 (closed strings with massless right and left moving spinors having the same chirality) by some kind of low energy limit. \\
 The conjecture was rephrased by Witten \cite{witt2} in proposing that the correlation (Schwin\-ger) functions  of the
 superconformal Yang-Mills theory may be obtained as asymptotic (boundary) limits of  5-dimenional supergravity on
$AdS_5$ plus Kaluza-Klein modes related to the compact $S^5,$ by means of the associated generating partition 
functions for the supergravity and the gauge theory, respectively. The dimensions of operators in the
superconformal gauge theory could be determined from masses in the $AdS_5$ supergravity theory (for vanishing masses those
 dimensions become ``canonical''). An important point is that strongly coupled
Yang-Mills theories correspond to weakly coupled supergravity, thus allowing -- in principle -- to calculate strong
 coupling effects in the gauge sector by perturbation theory in the corresponding  supergravity sector. 

 The hypothesis has caused a
lot of excitement, with (partial) confirmations for special cases or models and also by using the conjecture as a
working hypothesis for the analysis of certain problems, e.g.\ strong-coupling problems in gauge theories.

 I am unable to do justice to the many works and people working in the
field and I refer to reviews for further insights \cite{malrev}. 

The conjecture as phrased by Witten suggests the question whether such or a similar correspondence can perhaps be achieved
without refering to superstrings. This question was investigated with some success by Rehren in the framework of algebraic
 quantum field theory \cite{reh1}. A recent summary of results can be found in the Thesis \cite{rehrev}.
\subsubsection{``Unparticles''}
Physical systems with a discrete mass spectrum like the standard model of elementary particles cannot be scale and
 conformally invariant. Last year Georgi suggested that nevertheless  conformally covariant operators with definite
 (dynamical)
 scale dimension $d_U$ from an independent conformally invariant field theory might couple to standard model operators:
At suffiently high energies this could lead to a ``non-standard'' loss of energy and momentum in the form of ``unparticles''
 in very high energy
 reactions of standard model particles \cite{geor}. Though still extremely speculative this might lead to another
 interesting application of conformal symmetries, at least theoretically!
 
\section*{Acknowledgements}
\addcontentsline{toc}{section}{\protect\numberline{}{Acknowledgements}}
 I take the opportunity to thank a number of those people who in my early academic years helped
me to understand the possible physical role of the conformal group, by encouragements, criticisms or other helpful supports:
F.\ Bopp, W.\ Heisenberg, J.\ Wess, G.\ H\"{o}hler, K.\ Dietz, E.P.\ Wigner, J.\ Cornwall, Y.\ Nambu, H.\ Leutwyler,
 G.\ Mack, K.G.\ Wilson, M.\ Gell-Mann, R.\ Gatto, R.\ Jackiw, J.D.\ Bjorken, A.\ Salam, B.\ Renner and others.

 I thank the Libraries of DESY, of the University of Hamburg
and  that of its Mathematics Department and finally the Library of the University of G\"{o}ttingen for their continuous
 help in securing the rather large selection of literature for the present paper.
 
 Again I thank the DESY Theory Group for its very
kind hospitality after my retirement from the Institute for Theoretical Physics of the RWTH Aachen. 

I thank the Editor
F.W.\ Hehl of the Annalen der Physik for asking me to contribute to the Minkowski Centennial Issue, because without that
request this paper probably would not have been written.

I am very much indebted to my son David and to T.M.\ Trzeciak for creating the figures.

 Again most of my thanks go to my wife Dorothea who once more had to endure my long outer and inner
 absences while I was preoccupied with writing this article!
\addcontentsline{toc}{section}{\protect\numberline{}{References}}


\begin{thebibliography}{300}
\bibitem{harr4} According to Taylor, Saddler and especially Pepper, Ref.\ \cite{harr2}, Harriot in 1594 used numerical
 tables from an
earlier book by Clavius: Theodosii Tripolitae sphaericorum libri III, Romae, Basaus, 1586;
 whether and when Clavius' Astrolabium
from 1593 (or 1611) arrived in England and could have been seen by Harriot needs further investigation. Perhaps a book
written by a Jesuit was banned at that time in England!
\bibitem{aig3} On page 498: ... Secundum, ex contactu, quod \& Stereographice non incongru\`{e} potest appelari: quare
ut ea vox in usum venire liber\`{e} possit, dum alia melior non occurit, Lector, veniam dabis. ...
In his introduction to the special chapter on stereographic projections (``De Stereographice Altero Proiectionis Genere Ex
Oculi Contactu'') on the pages 572/73 he justifies the term in more detail and compares it, e.g.\ to ``Stereometria'' which
measures the capacity and extensions of bodies (``corporum dimensiones capacitatesque metitur'') [the Greek word ``stereos''
means ``rigid, solid, massive''].
\end{thebibliography}
\end{document}